\documentclass[aps,preprint]{revtex4}
\usepackage{amssymb}

\usepackage{amsmath}
\usepackage[dvips]{graphicx}

\input{tcilatex}

\begin{document}

\title{Oblique frozen modes in periodic layered media}
\author{A. Figotin and I. Vitebskiy}
\affiliation{Department of Mathematics, University of California at Irvine, CA 92697}

\begin{abstract}
We study the classical scattering problem of a plane electromagnetic wave
incident on the surface of semi-infinite periodic stratified media
incorporating anisotropic dielectric layers with special oblique orientation
of the anisotropy axes. We demonstrate that an obliquely incident light,
upon entering the periodic slab, gets converted into an abnormal grazing
mode with huge amplitude and zero normal component of the group velocity.
This mode cannot be represented as a superposition of extended and
evanescent contributions. Instead, it is related to a general (non-Bloch)
Floquet eigenmode with the amplitude diverging linearly with the distance
from the slab boundary. Remarkably, the slab reflectivity in such a
situation can be very low, which means an almost 100\% conversion of the
incident light into the \emph{axially frozen mode} with the electromagnetic
energy density exceeding that of the incident wave by several orders of
magnitude. The effect can be realized at any desirable frequency, including
optical and UV frequency range. The only essential physical requirement is
the presence of dielectric layers with proper oblique orientation of the
anisotropy axes. Some practical aspects of this phenomenon are considered.
\end{abstract}

\maketitle

\section{Introduction}

Electromagnetic properties of periodic stratified media have been a subject
of extensive research for decades (see, for example, \cite%
{Strat1,Strat2,Strat3} and references therein). Of particular interest has
been the case of periodic stacks ($1D$ photonic crystals) made up of
lossless dielectric components with different refractive indices. Photonic
crystals with one-dimensional periodicity had been widely used in optics
long before the term ''photonic crystals'' was invented.

\begin{figure}[tbph]
\includegraphics[scale=0.9, viewport= -100 0 400 220, clip]{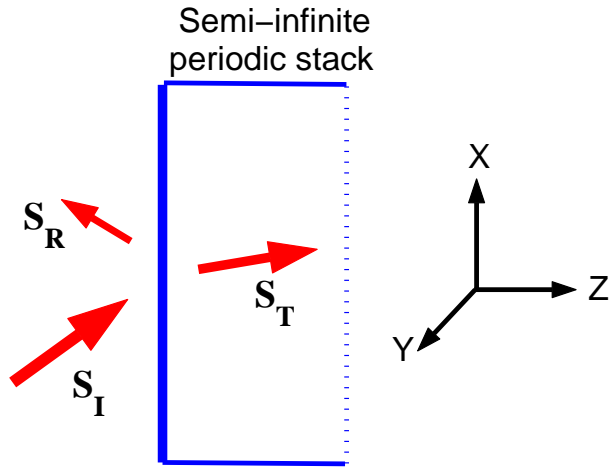}
\caption{The scattering problem for a semi-infinite periodic layered medium. 
$\vec{S}_{I}$, $\vec{S}_{R}$ and $\vec{S}_{T}$ are the energy density fluxes
of the incident, reflected and transmitted waves, respectively. The
transmitted wave $\Psi _{T}$ is a superposition of two Bloch eigenmodes,
each of which can be either extended or evanescent. Only extended modes can
transfer the energy in the $z$ direction.}
\label{SIS1}
\end{figure}

Let us look at the classical problem of a plane electromagnetic wave
incident on the surface of semi-infinite plane-parallel periodic array, as
shown in Fig. \ref{SIS1}. 
The well known effects of the slab periodicity are: (i) the possibility of
omnidirectional reflectance when the incident radiation is reflected by the
slab, regardless of the angle of incidence; (ii) the possibility of negative
refraction , when the tangential component of the energy flux $\vec{S}_{T}$
of transmitted wave is antiparallel to that of the incident wave; (iii)
dramatic slowdown of the transmitted wave near photonic band edge frequency,
where the normal component of the transmitted wave group velocity $\vec{u}$
vanishes along with the respective energy flux $\vec{S}_{T}$. The extensive
discussion on the subject and numerous references can be found in \cite%
{Omni1,Omni2,Omni3,
NegRefr1,NegRefr2,Joann1,Scalora1,Agua01,Scalora2,Scalora97}. All the above
effects can occur even in the simplest case of a semi-infinite periodic
array of two isotropic dielectric materials with different refractive
indices, for example, glass and air. The majority of known photonic crystals
fall into this category. The introduction of \emph{dielectric anisotropy},
however, can bring qualitatively new features to electromagnetic properties
of periodic stratified media and open up new opportunities for practical
applications (see, for example, a recent publication \cite{Manda03}). One of
such phenomena is the subject of this work.

\subsection{The Axially Frozen Mode (AFM)}

\begin{figure}[tbph]
\includegraphics[scale=0.9, viewport= -30 0 700 220, clip]{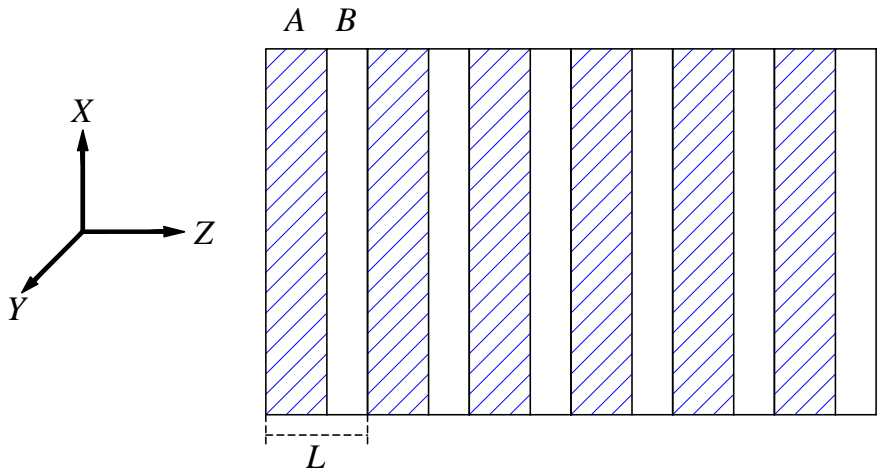}
\caption{Periodic layered structure with two layers $A$ and $B$ in a
primitive cell $L$. The $A$ layers (hatched) are anisotropic with one of the
principle axes of the dielectric permittivity tensor $\hat{\protect%
\varepsilon}$ making an oblique angle with the normal $z$ to the layers ($%
\protect\varepsilon _{xz}\neq 0$). The $B$ layers are isotropic. The $x-z$
plane coincides with the mirror plane $m_{y}$ of the stack.}
\label{StackAB}
\end{figure}

\begin{figure}[tbph]
\includegraphics[scale=0.9, viewport= -100 0 400 220, clip]{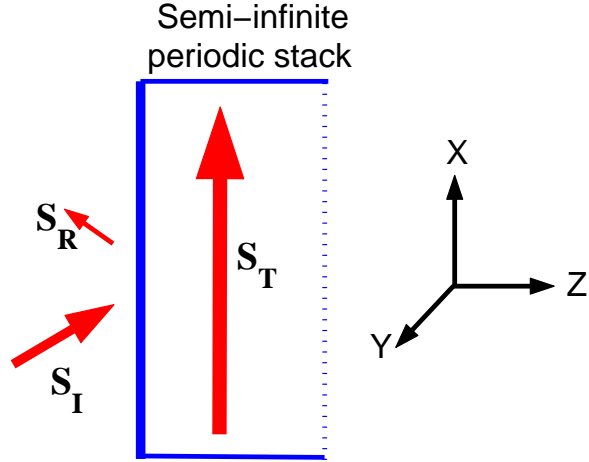}
\caption{An incident plane wave with unity energy density flux and certain
angle of incidence is converted into the AFM with huge amplitude, tangential
group velocity, and nearly tangential energy flux $\vec{S}_{T}$. The normal
components $\left( \vec{S}_{I}\right) _{z}$ and $\left( \vec{S}_{T}\right)
_{z}$ of the incident and transmitted waves energy flux are comparable in
magnitude.}
\label{SIS2}
\end{figure}

Consider a semi-infinite periodic stack with at least one of the
constituents being an anisotropic dielectric material with oblique
orientation of anisotropic axis. A simple example of such an array is
presented in Fig. \ref{StackAB}. 
We will show that under certain physical conditions, a monochromatic plane
wave incident on the semi-infinite slab is converted into an abnormal
electromagnetic mode with huge amplitude and nearly tangential energy
density flux, as illustrated in Fig. \ref{SIS2}. 
Such a wave will be referred to as the \emph{Axially Frozen Mode }(AFM). The
use of this term is justified because the normal (axial) component $u_{z}$
of the respective group velocity becomes vanishingly small, while the
amplitude of the AFM can exceed the amplitude of the incident plane wave by
several orders of magnitude.

\begin{figure}[tbph]
\includegraphics[scale=0.9, viewport= 0 0 700 220, clip]{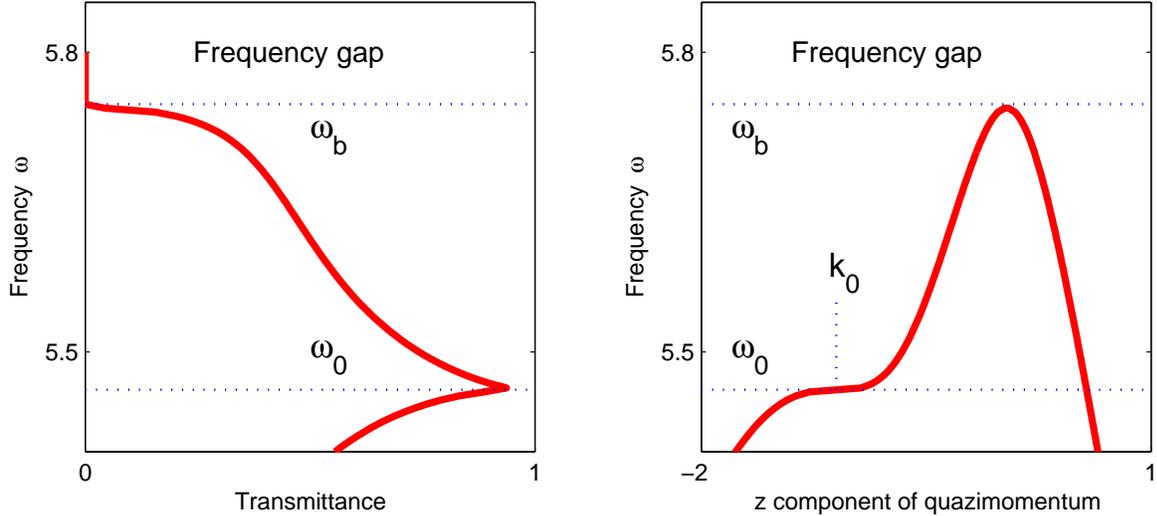}
\caption{(a) The transmittance $\protect\tau $ of periodic semi-infinite
slab vs. frequency at fixed direction $\vec{n}$ of the incidence. At the
frequency $\protect\omega _{0}$ of the AFM, $\protect\tau $ is close to
unity, which implies that the incident wave almost completely gets converted
into the AFM. (b) The respective axial dispersion relation $\protect\omega %
\left( k_{z}\right) $ at fixed $\left( n_{x},n_{y}\right) $ from Eq. (12).
At $k_{z}=k_{0}$ and $\protect\omega =\protect\omega _{0}$ this spectral
branch develops a stationary inflection point (16) associated with the AFM
regime. $\protect\omega _{b}$ is the edge of the frequency band for a given $%
\left( n_{x},n_{y}\right) $. The values of $\protect\omega $ and $k$ are
expressed in units of $c/L$ and $1/L$, respectively.}
\label{tE}
\end{figure}

The group velocity $\vec{u}$ of the AFM is parallel to the semi-infinite
slab boundary and, therefore, the magnitude of the tangential component $%
\left( \vec{S}_{T}\right) _{\bot }$ of the respective energy density flux $%
\vec{S}_{T}$ is overwhelmingly larger than the magnitude of the normal
component $\left( \vec{S}_{T}\right) _{z}$. But, although $\left( \vec{S}%
_{T}\right) _{z}\ll \left( \vec{S}_{T}\right) _{\bot }$, the normal
component $\left( \vec{S}_{T}\right) _{z}$ of the energy density flux inside
the slab is still comparable with that of the incident plane wave in vacuum.
This property persists even if the normal component $u_{z}$ of the wave
group velocity inside the slab vanishes, i.e.,%
\begin{equation}
\left( \vec{S}_{T}\right) _{z}>0\text{,}\;\text{if}\;u_{z}=0.  \label{Sz > 0}
\end{equation}%
The qualitative explanation for this is that the infinitesimally small value
of $u_{z}$ is offset by huge magnitude of the energy density $W$ in the AFM.
As the result, the product $u_{z}W$, which determines the normal component $%
\left( \vec{S}_{T}\right) _{z}$ of the energy flux, remains finite. The
above behavior is totally different from what happens in the vicinity of a
photonic band edge, where the normal component $u_{z}$ of the wave group
velocity vanishes too. Indeed, let us introduce the transmittance $\left(
\tau \right) $ and the reflectance $\left( \rho \right) $ of a lossless
semi-infinite slab%
\begin{equation}
\tau =1-\rho =\frac{\left( \vec{S}_{T}\right) _{z}}{\left( \vec{S}%
_{I}\right) _{z}},\;\;\rho =-\frac{\left( \vec{S}_{R}\right) _{z}}{\left( 
\vec{S}_{I}\right) _{z}}.  \label{t_e,r_e}
\end{equation}%
In line with Eq. (\ref{Sz > 0}), in the AFM\ regime the transmittance $\tau $
remains significant and can be even close to unity, as shown in an example
in Fig. \ref{tE}(a). In other words, the incident plane wave enters the slab
with little reflectance, where it turns into an abnormal AFM with
infinitesimally small normal component of the group velocity, huge
amplitude, and huge tangential component of the energy density flux. By
contrast, in the vicinity of a photonic band edge (at frequencies near $%
\omega =\omega _{b}$ in Fig. \ref{tE}(a)), the transmittance of
semi-infinite slab always vanishes, along with the normal component $u_{z}$
of the wave group velocity.

It turns out that at a given frequency $\omega _{0}$ the AFM regime can
occur only for a special direction $\vec{n}_{0}$ of the incident plane wave
propagation%
\begin{equation}
\vec{n}_{0}=\vec{n}_{0}\left( \omega _{0}\right) .  \label{n0,w0}
\end{equation}%
This special direction of incidence always makes an oblique angle with the
normal $z$ to the layers. To find $\vec{n}_{0}$ for a given $\omega _{0}$
or, conversely, to find $\omega _{0}$ for a given $\vec{n}_{0}$, one has to
solve the Maxwell equations in the periodic stratified medium. This problem
will be addressed in Section 3. In Section 2 we consider the relation
between the AFM regime and the singularity of the electromagnetic dispersion
relation responsible for such a peculiar behavior. If the frequency $\omega $
and the direction of incidence $\vec{n}$ do not match explicitly as
prescribed by Eq. (\ref{n0,w0}), the AFM regime will be somewhat smeared.

\subsection{The vicinity of the AFM regime}

\begin{figure}[tbph]
\includegraphics[scale=0.9, viewport= 0 0 700 220, clip]{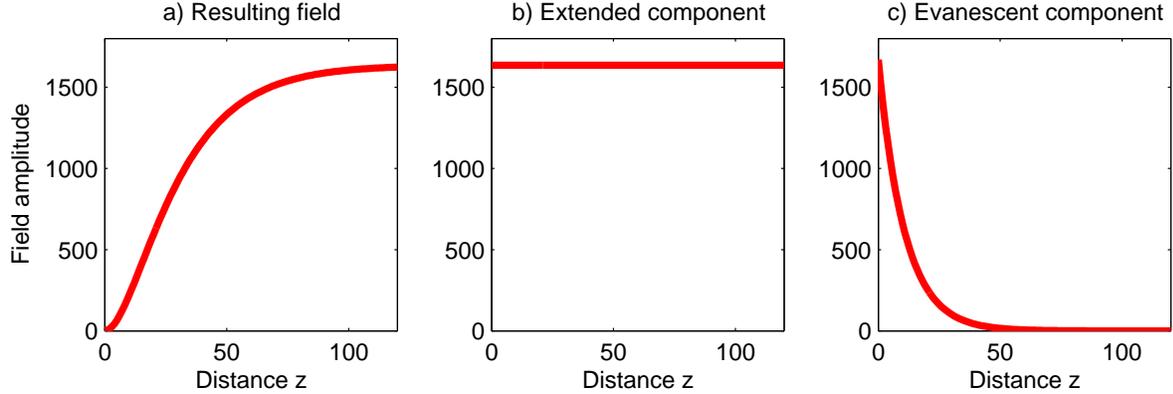}
\caption{Destructive interference of the extended and evanescent components
of the resulting electromagnetic field (4) inside semi-infinite slab in
close proximity of the AFM regime: (a) the amplitude $\left| \Psi _{T}\left(
z\right) \right| ^{2}$ of the resulting field, (b) the amplitude $\left|
\Psi _{ex}\left( z\right) \right| ^{2}$ of the extended contribution, (c)
the amplitude $\left| \Psi _{ev}\left( z\right) \right| ^{2}$ of the
evanescent contribution. The amplitude $\left| \Psi _{I}\right| ^{2}$ of the
incident wave is unity. The distance $z$ from the slab boundary is expressed
in units of $L$.}
\label{AMz}
\end{figure}

Let $\Psi _{T}\left( z\right) $ be the transmitted electromagnetic field
inside the semi-infinite slab (the explicit definition of $\Psi _{T}\left(
z\right) $ is given in Eqs. (\ref{ME4}) and (\ref{T=ex+ev})). It turns out
that in the vicinity of the AFM regime, $\Psi _{T}\left( z\right) $ is a
superposition of the extended and evanescent Bloch eigenmodes%
\begin{equation}
\Psi _{T}\left( z\right) =\Psi _{ex}\left( z\right) +\Psi _{ev}\left(
z\right) ,\;\;z>0,  \label{Psi Tz}
\end{equation}%
where $\Psi _{ex}\left( z\right) $ is an extended mode with $u_{z}>0$, and $%
\Psi _{ev}\left( z\right) $ is an evanescent mode with $\Im k_{z}>0$. As
shown in an example in Fig. \ref{AMz}, both the contributions to $\Psi
_{T}\left( z\right) $ have huge and nearly equal and opposite values near
the slab boundary, so that their superposition (\ref{Psi Tz}) at $z=0$ is
small enough to satisfy the boundary condition (\ref{BC}). As the distance $%
z $ from the slab boundary increases, the evanescent component $\Psi
_{ev}\left( z\right) $ decays exponentially, while the amplitude of the
extended component $\Psi _{ex}\left( z\right) $ remains constant and huge.
As the result, the field amplitude $\left| \Psi _{T}\left( z\right) \right|
^{2}$ reaches its huge saturation value $\left| \Psi _{ex}\right| ^{2}$ at a
certain distance from the slab boundary (see Eqs. (\ref{Psi ev}), (\ref%
{Psi(N)}) and (\ref{PsiT0})). 

When the direction of incidence $\vec{n}$ tends to its critical value $\vec{n%
}_{0}$ for a given frequency $\omega _{0}$, the respective saturation value $%
\left| \Psi _{ex}\right| ^{2}$ of the AFM amplitude $\left| \Psi _{T}\left(
z\right) \right| ^{2}$ diverges as $\left| \vec{n}-\vec{n}_{0}\right|
^{-2/3} $. Conversely, when the frequency $\omega $ tends to its critical
value $\omega _{0}$ for a given direction of incidence $\vec{n}_{0}$, the
saturation value of the AFM amplitude diverges as $\left| \omega -\omega
_{0}\right| ^{-2/3}$. In the real situation, of course, the AFM amplitude
will be limited by such physical factors as: (i) nonlinear effects, (ii)
electromagnetic losses, (iii) structural imperfections of the periodic
array, (iv) finiteness of the slab dimensions, (v) deviation of the incident
radiation from a perfect plane monochromatic wave.

\begin{figure}[tbph]
\includegraphics[scale=0.9, viewport= 0 0 400 400, clip]{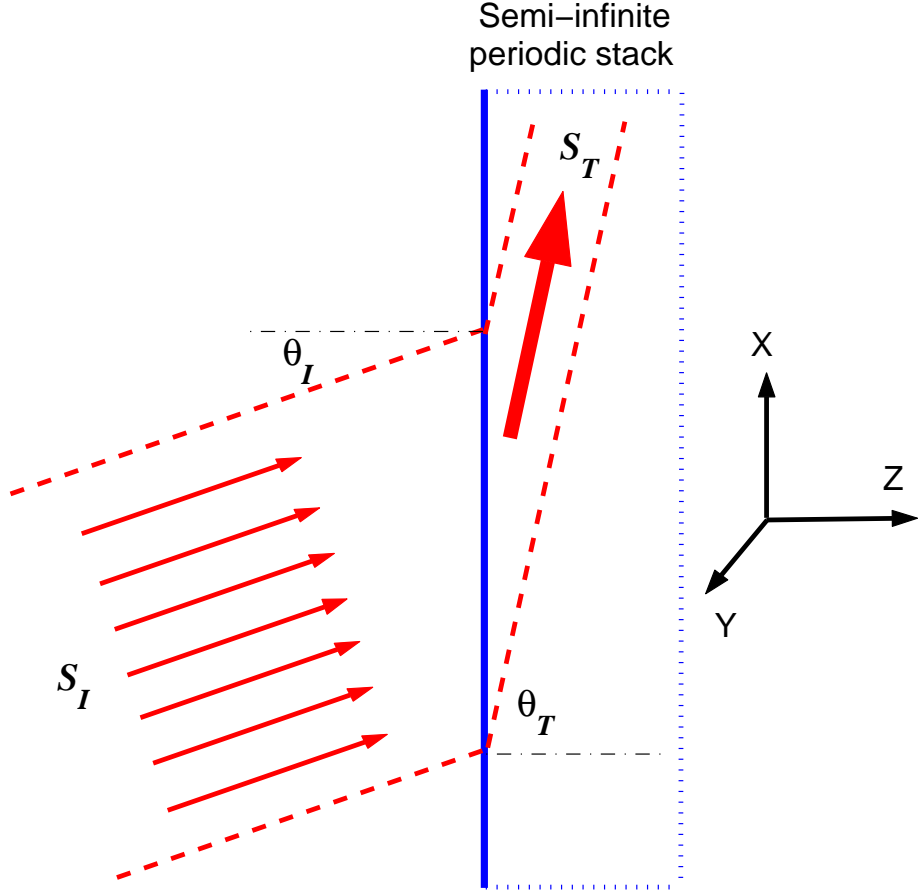}
\caption{Incident and transmitted (refracted) waves in the vicinity of the
AFM regime. The reflected wave is not shown. $\protect\theta _{I}$ and $%
\protect\theta _{T}$ are the incidence and refraction angles, $S_{I}$ and $%
S_{T}$ are the energy density fluxes of the incident and transmitted waves.
Both the energy density and the energy density flux in the transmitted wave
are much larger than the respective values in the incident wave. However,
the total power transmitted by the refracted wave is smaller by factor $%
\protect\tau $, due to much smaller cross-section area of the nearly grazing
transmitted wave.}
\label{Beam}
\end{figure}

Fig. \ref{Beam} gives a good qualitative picture of what really happens in
the vicinity of the AFM regime. 
Consider a wide monochromatic beam of frequency $\omega $ incident on the
surface of semi-infinite photonic slab. The direction of incidence $\vec{n}%
_{0}\parallel \vec{S}_{I}$ is chosen so that the condition (\ref{n0,w0}) of
the AFM regime is satisfied at $\omega =\omega _{0}$. As frequency $\omega $
tends to $\omega _{0}$ from either direction, the normal component $u_{z}$
of the transmitted wave group velocity approaches zero, while the tangential
component $\vec{u}_{\perp }$ remains finite 
\begin{equation}
u_{z}\thicksim \left| \omega -\omega _{0}\right| ^{2/3}\rightarrow 0,\;\vec{u%
}_{\perp }\rightarrow \vec{u}_{0}\;\text{\ \ as \ }\omega \rightarrow \omega
_{0}.  \label{u_z, u_n}
\end{equation}%
This relation together with the equality%
\begin{equation}
\frac{\pi }{2}-\theta _{T}=\arctan \frac{u_{z}}{u_{\perp }}  \label{theta_T}
\end{equation}%
involving the refraction angle $\theta _{T}$, yield%
\begin{equation}
\frac{\pi }{2}-\theta _{T}\thicksim \left| \omega -\omega _{0}\right|
^{2/3}\rightarrow 0\;\text{\ \ as \ }\omega \rightarrow \omega _{0}.
\label{theta 1}
\end{equation}%
Hence, in the vicinity of the AFM regime, the transmitted (refracted)
electromagnetic wave can be viewed as a \emph{grazing mode}. The most
important and unique feature of this grazing mode directly relates to the
fact that the transmittance $\tau $ of the semi-infinite slab remains finite
even at $\omega =\omega _{0}$ (see, for example, Fig. \ref{tE}(a)). Indeed,
let $A_{I}$ and $A_{T}$ be the cross-section areas of the incident and
transmitted (refracted) beams, respectively. Obliviously,%
\begin{equation}
\frac{A_{T}}{A_{I}}=\frac{\cos \theta _{T}}{\cos \theta _{I}}
\label{theta 2}
\end{equation}%
Let us also introduce the quantities 
\begin{equation}
U_{I}=A_{I}S_{I},\;U_{T}=A_{T}S_{T},  \label{theta 3}
\end{equation}%
where $S_{I}$ and $S_{T}$ are the energy density fluxes of the incident and
transmitted waves. $U_{I}$ and $U_{T}$ are the total power transmitted by
the incident and transmitted (refracted) beams, respectively. The
expressions (\ref{theta 2}) and (\ref{theta 3}) imply that%
\begin{equation}
\frac{U_{T}}{U_{I}}=\frac{S_{T}\cos \theta _{T}}{S_{I}\cos \theta _{I}}=%
\frac{\left( S_{T}\right) _{z}}{\left( S_{I}\right) _{z}}=\tau
\label{theta 4}
\end{equation}%
which is nothing more than a manifestation of the energy conservation law.
Finally, Eq. (\ref{theta 4}), together with the formula (\ref{theta 1}),
yield%
\begin{equation}
S_{T}=\tau S_{I}\frac{\cos \theta _{I}}{\cos \theta _{T}}\thicksim \left|
\omega -\omega _{0}\right| ^{-2/3}\rightarrow \infty \;\text{\ \ as \ }%
\omega \rightarrow \omega _{0}.  \label{theta 5}
\end{equation}%
where we have taken into account that $\tau S_{I}\cos \theta _{I}$ is
limited (of the order of magnitude of unity) as \ $\omega \rightarrow \omega
_{0}$. By contrast, in the vicinity of the photonic band edge the
transmittance $\tau $ of the semi-infinite slab vanishes along with the
energy density flux $S_{T}$ of the transmitted (refracted) wave.

The expressions (\ref{theta 1}) and (\ref{theta 5}) show that in the
vicinity of the AFM regime, the transmitted wave behaves like a grazing mode
with huge and nearly tangential energy density flux $S_{T}$ and very small
(compared to that of the incident beam) cross-section area $A_{T}$, so that
the total power $U_{T}=A_{T}S_{T}$ associated with the transmitted wave
cannot exceed the total power $U_{I}$ of the incident wave: $U_{T}=\tau
U_{I}\leq U_{I}$.

The above qualitative consideration is only valid on the scales exceeding
the size $L$ of the unit cell (which is of the order of magnitude of $%
c/\omega $) and more importantly, exceeding the \emph{transitional distance} 
$l=\left( \Im k_{ev}\right) ^{-1}$ from the slab boundary where the
evanescent mode contribution to the resulting electromagnetic field $\Psi
_{T}\left( z\right) $ is still significant. The latter means that the width
of both the incident and the refracted beams must be much larger than $l$.
If the above condition is not met, we cannot treat the transmitted wave as a
beam, and the expressions (\ref{theta 1}) through (\ref{theta 5}) do not
apply. Instead, we would have to use the explicit electrodynamic expressions
for $\Psi _{T}\left( z\right) $, such as the asymptotic formula (\ref{PsiT0}%
). Note that if the direction $\vec{n}$ of the incident wave propagation and
the frequency $\omega $ \emph{exactly match} the condition (\ref{n0,w0}) for
the AFM regime, the transmitted wave $\Psi _{T}\left( z\right) $ does not
reduce to a superposition (\ref{Psi Tz}) of canonical Bloch eigenmodes.
Instead, the AFM is described by a general Floquet eigenmode $\Psi
_{01}\left( z\right) $ from Eq. (\ref{Psi 01}), which diverges inside the
slab as $z$, until the nonlinear effects or other limiting factors come into
play. The related mathematical analysis is provided in Sections 3 and 4.

In some respects, the remarkable behavior of the AFM, is similar to that of
the \emph{frozen mode} related to the phenomenon of \emph{electromagnetic
unidirectionality }in nonreciprocal magnetic photonic crystals \cite%
{PRE01,PRB03}. In a unidirectional photonic crystal, electromagnetic
radiation of a certain frequency $\omega _{0}$ can propagate with finite
group velocity $\vec{u}\Vert z$ only in one of the two opposite directions,
say, from right to left. The problem with the electromagnetic
unidirectionality, though, is that it essentially requires the presence of
magnetic materials with strong circular birefringence (Faraday rotation) and
low losses at the frequency range of interest. Such materials are readily
available at the microwave frequencies, but at the infrared and optical
frequency ranges, finding appropriate magnetic materials is highly
problematic. Thus, at frequencies above $10^{12}$ Hz, the electromagnetic
unidirectionality along with the respective nonreciprocal magnetic mechanism
of the frozen mode formation may prove to be impractical. \emph{By contrast,
the occurrence of AFM does not require the presence of magnetic or any other
essentially dispersive components in the periodic stack. Therefore, the AFM
regime can be realized at any frequencies, including the infrared, optical,
and even ultraviolet frequency ranges}. The only essential physical
requirement is the presence of anisotropic dielectric layers with proper
orientation of the anisotropy axes. An example of such an array is shown in
Fig. \ref{StackAB}.

In Section 2 we establish the relation between the phenomenon of AFM and the
electromagnetic dispersion relation of the periodic layered medium. This
allows us to formulate strict and simple symmetry conditions for such a
phenomenon to occur, as well as to find out what kind of periodic stratified
media can exhibit the effect. Relevant theoretical analysis based on the
Maxwell equations in stratified media is carried out in Sections 3 and 4.
Finally, in Section 5 we discuss some practical aspects of the phenomenon.

\section{Dispersion relation with the AFM}

Now we establish the connection between the phenomenon of AFM and the
electromagnetic dispersion relation $\omega \left( \vec{k}\right) ,\;\vec{k}%
=(k_{x},k_{y},k_{z})$ of the periodic stratified medium. In a plane-parallel
stratified slab, the tangential components $(k_{x},k_{y})$ of the Bloch wave
vector $\vec{k}$ always coincide with those of the incident plane wave in
Figs. \ref{SIS1}, \ref{SIS2}, and \ref{Beam} while the normal component $%
k_{z}$ is different from that of the incident wave. To avoid confusion, in
further consideration, the $z$ component of the Bloch wave vector $\vec{k}$
inside the periodic slab will be denoted as $k$ without the subscript $z$,
namely%
\begin{equation*}
\text{Inside periodic stack: \ \ }\vec{k}=(k_{x},k_{y},k).
\end{equation*}%
The value of $k$ is found by solving the Maxwell equations in the periodic
stratified medium for given $\omega $ and $(k_{x},k_{y})$; $k$ is defined up
to a multiple of $2\pi /L$, where $L$ is the period of the layered structure.

\begin{figure}[tbph]
\includegraphics[scale=0.9, viewport= 0 0 700 500, clip]{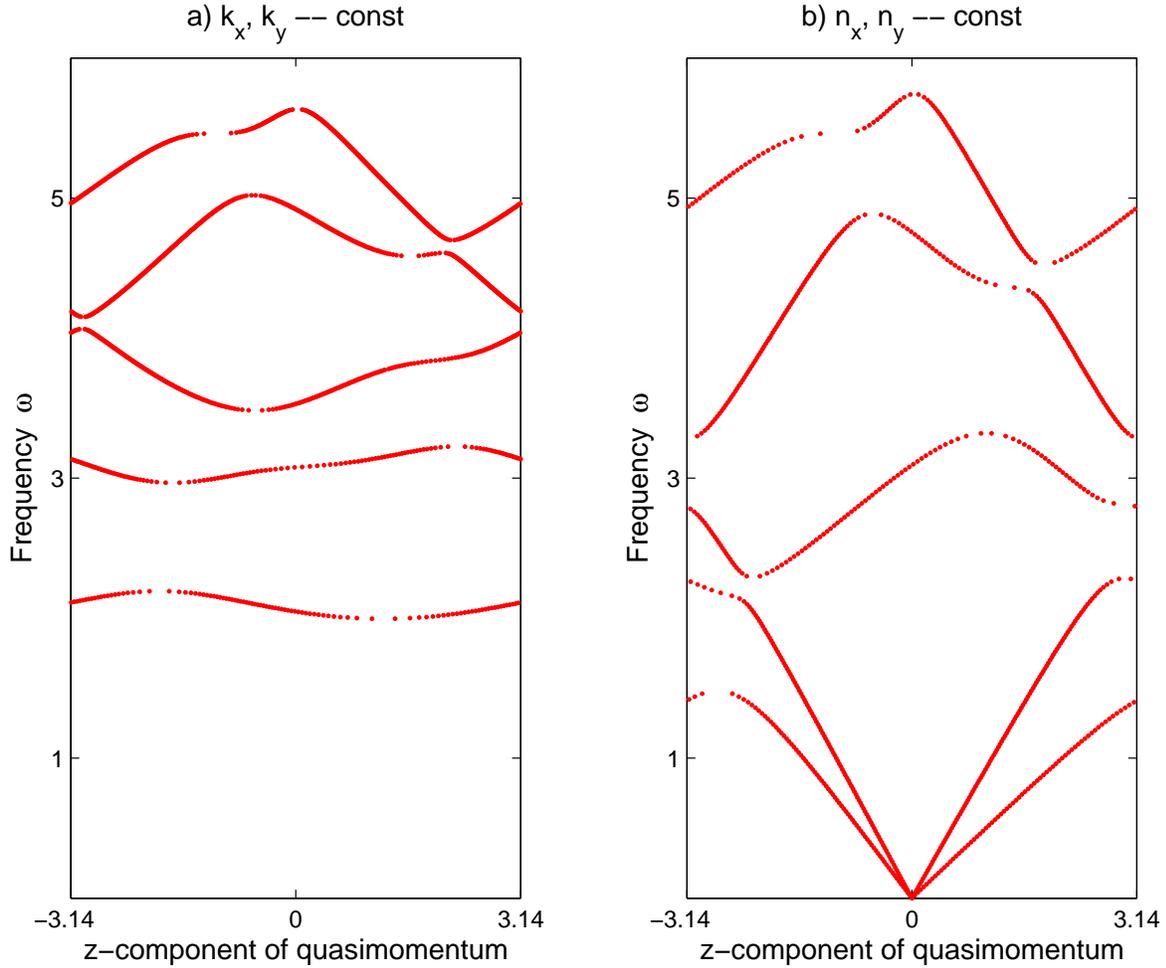}
\caption{The axial dispersion relation of anisotropic periodic stack in Fig.
2: \ (a) $\protect\omega (k_{z})$ for fixed values $(k_{x},k_{y})$ of the
tangential components of quasimomentum $\vec{k}$; (b) $\protect\omega %
(k_{z}) $ for fixed values $(n_{x},n_{y})$, defining the direction of
incidence. In the case of normal incidence, there would be no difference
between (a) and (b).}
\label{DRnk2}
\end{figure}

Consider now the frequency $\omega $ as function of $k$ for fixed $%
(k_{x},k_{y})$. A typical example of such a dependence is shown in Fig. \ref%
{DRnk2}(a). A large gap at the lowest frequencies is determined by the value
of the fixed tangential components $(k_{x},k_{y})$ of the quasimomentum $%
\vec{k}$. This gap vanishes in the case of normal incidence, when $%
k_{x}=k_{y}=0$. An alternative and more convenient representation for the
dispersion relation is presented in Fig. \ref{DRnk2}(b), where the plot of $%
\omega (k)$ is obtained for fixed $(n_{x},n_{y})$ based on%
\begin{equation}
(n_{x},n_{y})=(ck_{x}/\omega ,ck_{y}/\omega ).  \label{n(k)}
\end{equation}%
The pair of values $(n_{x},n_{y})$ coincide with the tangential components
of the unit vector $\vec{n}$ defining the direction of the incident plane
wave propagation. The dependence $\omega (k)$ for fixed $(n_{x},n_{y})$ or
for fixed $(k_{x},k_{y})$ will be referred to as the \emph{axial dispersion
relation}.

Suppose that for $\vec{k}=\vec{k}_{0}\;$and $\omega =\omega _{0}=\omega
\left( \vec{k}_{0}\right) $, one of the spectral branches $\omega \left(
k\right) $ develops a stationary inflection point for given $%
(k_{x},k_{y})=(k_{0x},k_{0y})$, i.e., 
\begin{equation}
\left. \left( \frac{\partial \omega }{\partial k}\right)
_{k_{x},k_{y}}\right| _{\vec{k}=\vec{k}_{0}}=0;\;\left. \left( \frac{%
\partial ^{2}\omega }{\partial k^{2}}\right) _{k_{x},k_{y}}\right| _{\vec{k}=%
\vec{k}_{0}}=0;\;\left. \left( \frac{\partial ^{3}\omega }{\partial k^{3}}%
\right) _{k_{x},k_{y}}\right| _{\vec{k}=\vec{k}_{0}}\neq 0,  \label{kInfl}
\end{equation}%
The value%
\begin{equation}
u_{z}=\left( \frac{\partial \omega }{\partial k}\right) _{k_{x},k_{y}}
\label{u_z}
\end{equation}%
in Eq. (\ref{kInfl}) is the axial component of the group velocity, which
vanishes at $\vec{k}=\vec{k}_{0}$. Observe that%
\begin{equation}
u_{x}=\left( \frac{\partial \omega }{\partial k_{x}}\right) _{k,k_{y}}\text{
\ and \ }u_{y}=\left( \frac{\partial \omega }{\partial k_{y}}%
\right)_{k,k_{x}},  \label{u_x, u_y}
\end{equation}%
representing the tangential components of the group velocity, may not be
zeros at $\vec{k}=\vec{k}_{0}$.

Notice that instead of (\ref{kInfl}), one can use another definition of the
stationary inflection point%
\begin{equation}
\left. \left( \frac{\partial \omega }{\partial k}\right)
_{n_{x},n_{y}}\right| _{\vec{k}=\vec{k}_{0}}=0,\;\left. \left( \frac{%
\partial ^{2}\omega }{\partial k^{2}}\right) _{n_{x},n_{y}}\right| _{\vec{k}=%
\vec{k}_{0}}=0,\;\left. \left( \frac{\partial ^{3}\omega }{\partial k^{3}}%
\right) _{n_{x},n_{y}}\right| _{\vec{k}=\vec{k}_{0}}\neq 0.  \label{nInfl}
\end{equation}%
The partial derivatives in Eqs. (\ref{nInfl}) are taken at constant $%
(n_{x},n_{y})$,$\ $rather than at constant $(k_{x},k_{y})$. Observe that the
definitions (\ref{kInfl}) and (\ref{nInfl}) are equivalent, and we will use
both of them.

In Fig. \ref{tE}(b) we reproduced an enlarged fragment of the upper spectral
branch of the axial dispersion relation in Fig. \ref{DRnk2}(b). For the
chosen $(n_{x},n_{y})$, this branch develops a stationary inflection point (%
\ref{nInfl}) at $\omega =\omega _{0}$ and $k=k_{0}$. The extended Bloch
eigenmode with $\omega =\omega _{0}$ and $\vec{k}=\vec{k}_{0}$, associated
with the stationary inflection point, turns out to be directly related to
the \emph{axially frozen mode} (AFM).

In Sections 3 and 4, based on the Maxwell equations, we prove that the
singularity (\ref{nInfl}) (or, equivalently, (\ref{kInfl})) indeed leads to
the very distinct AFM\ regime in the semi-infinite periodic stack. We also
show that a necessary condition for such a singularity and, therefore, a
necessary condition for the AFM existence is the following property of the
axial dispersion relation of the periodic stack%
\begin{equation}
\omega \left( k_{x},k_{y},k\right) \neq \omega \left( k_{x},k_{y},-k\right) 
\text{ \ or, equivalently,\ \ }\omega \left( n_{x},n_{y},k\right) \neq
\omega \left( n_{x},n_{y},-k\right)  \label{asym}
\end{equation}%
This property will be referred to as the \emph{axial spectral asymmetry}.
Evidently, the axial dispersion relations presented in Fig. \ref{DRnk2},
satisfy this criterion. Leaving the proof of the above statements to Section
3, let us look at the constraints imposed by the criterion (\ref{asym}) on
the geometry and composition of the periodic stack.

\subsection{Conditions for the axial spectral asymmetry}

First of all, notice that a periodic array would definitely have an \emph{%
axially symmetric} dispersion relation%
\begin{equation}
\omega \left( k_{x},k_{y},k\right) =\omega \left( k_{x},k_{y},-k\right) 
\text{\ \ or, equivalently,\ \ }\omega \left( n_{x},n_{y},k\right) =\omega
\left( n_{x},n_{y},-k\right)  \label{sym z}
\end{equation}%
if the symmetry group $G$ of the periodic stratified medium includes any of
the following two symmetry operations%
\begin{equation}
m_{z},\;2_{z}^{\prime }=2_{z}\times R,  \label{2'_z}
\end{equation}%
where $m_{z}$ is the mirror plane parallel to the layers, $2_{z}$ is the
2-fold rotation about the $z$ axis, and $R$ is the time reversal operation.
Indeed, since $2_{z}\left( k_{x},k_{y},k\right) =\left(
-k_{x},-k_{y},k\right) $ and $R\left( k_{x},k_{y},k\right) =\left(
-k_{x},-k_{y},-k\right) $, we have%
\begin{equation*}
2_{z}^{\prime }\left( k_{x},k_{y},k\right) =\left( k_{x},k_{y},-k\right) ,
\end{equation*}%
which implies the relation (\ref{sym z}) for arbitrary $\left(
k_{x},k_{y}\right) $. The same is true for the mirror plane $m_{z}$%
\begin{equation*}
m_{z}\left( k_{x},k_{y},k\right) =\left( k_{x},k_{y},-k\right) .
\end{equation*}%
Consequently, a necessary condition for the axial spectral asymmetry (\ref%
{asym}) of a periodic stack is the absence of the symmetry operations (\ref%
{2'_z}), i.e.,%
\begin{equation}
m_{z}\notin G\text{ \ and \ }2_{z}^{\prime }\notin G.  \label{asymC1}
\end{equation}%
In reciprocal (nonmagnetic) media, where by definition, $R\in G$, instead of
Eq. (\ref{asymC1}) one can use the following requirement%
\begin{equation}
m_{z}\notin G\text{ \ and \ }2_{z}\notin G.  \label{asymC2}
\end{equation}

Note, that the \emph{axial spectral symmetry} (\ref{sym z}) is different
from the \emph{bulk spectral symmetry}%
\begin{equation}
\omega \left( k_{x},k_{y},k\right) =\omega \left( -k_{x},-k_{y},-k\right)
\label{symm B}
\end{equation}%
For example, the space inversion $I$ and/or the time reversal $R$, if
present in $G$, ensure the bulk spectral symmetry (\ref{symm B}), but
neither $I$ nor $R$ ensures the axial spectral symmetry (\ref{sym z}).

\subsubsection{Application of the criterion (\ref{asymC2}) to deferent
periodic stacks.}

The condition (\ref{asymC2}) for the axial spectral asymmetry imposes
certain restrictions on the geometry and composition of the periodic
stratified medium, as well as on the direction of the incident wave
propagation.

\paragraph{Restrictions on the geometry and composition of periodic stack.}

First of all, observe that a common periodic stack made up of \emph{isotropic%
} dielectric components with different refractive indices, always has
axially symmetric dispersion relation (\ref{sym z}), no matter how
complicated the periodic array is or how many different isotropic materials
are involved. To prove this, it suffices to note that such a stack always
supports the symmetry operation $2_{z}$.

In fact, the symmetry operation $2_{z}$ holds in the more general case when
all the layers are either isotropic, or have a purely \emph{in-plane }%
anisotropy%
\begin{equation}
\mathbf{\hat{\varepsilon}}=\left[ 
\begin{array}{ccc}
\varepsilon _{xx} & \varepsilon _{xy} & 0 \\ 
\varepsilon _{xy} & \varepsilon _{yy} & 0 \\ 
0 & 0 & \varepsilon _{zz}%
\end{array}%
\right]  \label{e_xy = e_yx}
\end{equation}%
Obviously, the in-plane anisotropy (\ref{e_xy = e_yx}) does not remove the
symmetry operation $2_{z}$ and, therefore, the property (\ref{sym z}) of the
axial spectral symmetry holds in this case. Thus, we can state that in order
to display the axial spectral asymmetry, the periodic stack must include at
least one anisotropic component, either uniaxial or biaxial. In addition,
one of the principle axes of the respective dielectric permittivity tensor $%
\hat{\varepsilon}$ must make an oblique angle with the normal to the layers,
which means that \emph{at least} \emph{one of the two components }$%
\varepsilon _{xz}$\emph{\ and }$\varepsilon _{yz}$\emph{\ of the respective
dielectric tensor must be nonzero}.

\begin{figure}[tbph]
\includegraphics[scale=0.9, viewport= 0 0 700 220, clip]{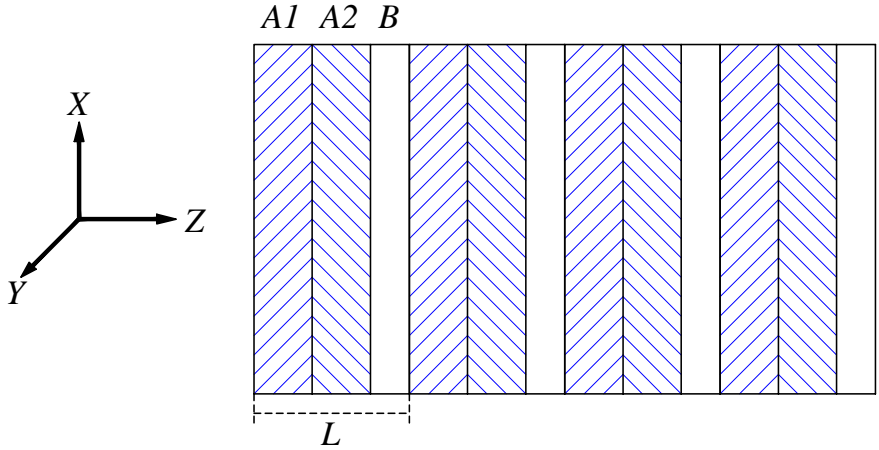}
\caption{Periodic stack composed of anisotropic layers $A1$ and $A2$, which
are the mirror images of each other, and isotropic layers $B$. This stack
has axially symmetric dispersion relation and does not support the AFM
regime. This is true even if the $B$ layers are removed.}
\label{StackAAB}
\end{figure}

The above requirement gives us a simple and useful idea on what kind of
periodic stratified media can support the axial spectral asymmetry and the
AFM regime. But this is not a substitute for the stronger symmetry criterion
(\ref{asymC1}) or (\ref{asymC2}). For example, although the periodic stack
in Fig. \ref{StackAAB} 
includes the $A$ layers identical to those in Fig. \ref{StackAB}, this stack
does not meet the criterion (\ref{asymC1}) for the axial spectral asymmetry.
Indeed, the stack in Fig. \ref{StackAAB} supports the mirror plane $m_{z}$,
which, according to the expression (\ref{2'_z}), ensures the axial spectral
symmetry.

\paragraph{Restriction on the direction of incident wave propagation\protect%
\medskip}

Consider now an important particular case $k_{x}=k_{y}=0$ of the normal
incidence. The criterion (\ref{asym}) reduces now to the simple requirement%
\begin{equation}
\omega \left( \vec{k}\right) \neq \omega \left( -\vec{k}\right) ,\;\text{%
where \ }\vec{k}=\left( 0,0,k\right) ,  \label{nrs}
\end{equation}%
of the bulk spectral asymmetry, which is prohibited in nonmagnetic photonic
crystals due to the time reversal symmetry. Therefore, in the nonmagnetic
case, we have the following additional condition for the axial spectral
asymmetry%
\begin{equation}
k_{\perp }=\sqrt{k_{x}^{2}+k_{y}^{2}}\neq 0,  \label{kxKy<>0}
\end{equation}%
implying that the AFM cannot be excited in a nonmagnetic semi-infinite stack
by a normally incident plane wave, i.e., \emph{the incident angle must be
oblique}.

Conditions (\ref{asymC2}) and (\ref{kxKy<>0}) may not be necessary in the
case of nonreciprocal magnetic stacks (see the details in \cite{PRE01}). But
as we mentioned earlier, at frequencies above $10^{12}$ Hz, the
nonreciprocal effects in common nonconducting materials are negligible.
Therefore, in order to have a robust AFM\ regime in the infrared or optical
frequency range, we must satisfy both requirements (\ref{asymC2}) and (\ref%
{kxKy<>0}), regardless of whether or not nonreciprocal magnetic materials
are involved.

As soon as the above conditions are met, one can always achieve the AFM
regime at any desirable frequency $\omega $ within certain frequency range $%
\Delta \omega $. The frequency range $\Delta \omega $ is determined by the
stack geometry and the dielectric materials used, while a specific value of $%
\omega $ within the range can be selected by the direction $\vec{n}$ of the
light incidence.

\subsection{Periodic stack with two layers in unit cell}

The simplest and the most practical example of a periodic stack supporting
the axial spectral asymmetry (\ref{asym}) and, thereby, the AFM regime, is
shown in Fig. \ref{StackAB}. It is made up of anisotropic $A$ layers
alternated with isotropic $B$ layers. The respective dielectric permittivity
tensors are 
\begin{equation}
\hat{\varepsilon}_{A}=\left[ 
\begin{array}{ccc}
\varepsilon _{xx} & 0 & \varepsilon _{xz} \\ 
0 & \varepsilon _{yy} & 0 \\ 
\varepsilon _{xz} & 0 & \varepsilon _{zz}%
\end{array}%
\right] ,\;\hat{\varepsilon}_{B}=\left[ 
\begin{array}{ccc}
\varepsilon _{B} & 0 & 0 \\ 
0 & \varepsilon _{B} & 0 \\ 
0 & 0 & \varepsilon _{B}%
\end{array}%
\right] .  \label{epsilon}
\end{equation}%
For simplicity, we assume%
\begin{equation}
\hat{\mu}_{A}=\hat{\mu}_{B}=\hat{I}.  \label{mu}
\end{equation}%
The stack in Fig. \ref{StackAB} has the monoclinic symmetry%
\begin{equation}
2_{y}/m_{y}  \label{2/m}
\end{equation}%
with the mirror plane $m_{y}$ normal to the $y$ - axis. Such a symmetry is
compatible with the necessary condition (\ref{asymC2}) for the AFM
existence. But as we will see below, the symmetry (\ref{2/m}) imposes
additional constraints on the direction $\vec{n}$ of the incident wave
propagation.

\begin{figure}[tbph]
\includegraphics[scale=0.9, viewport= 0 0 700 500, clip]{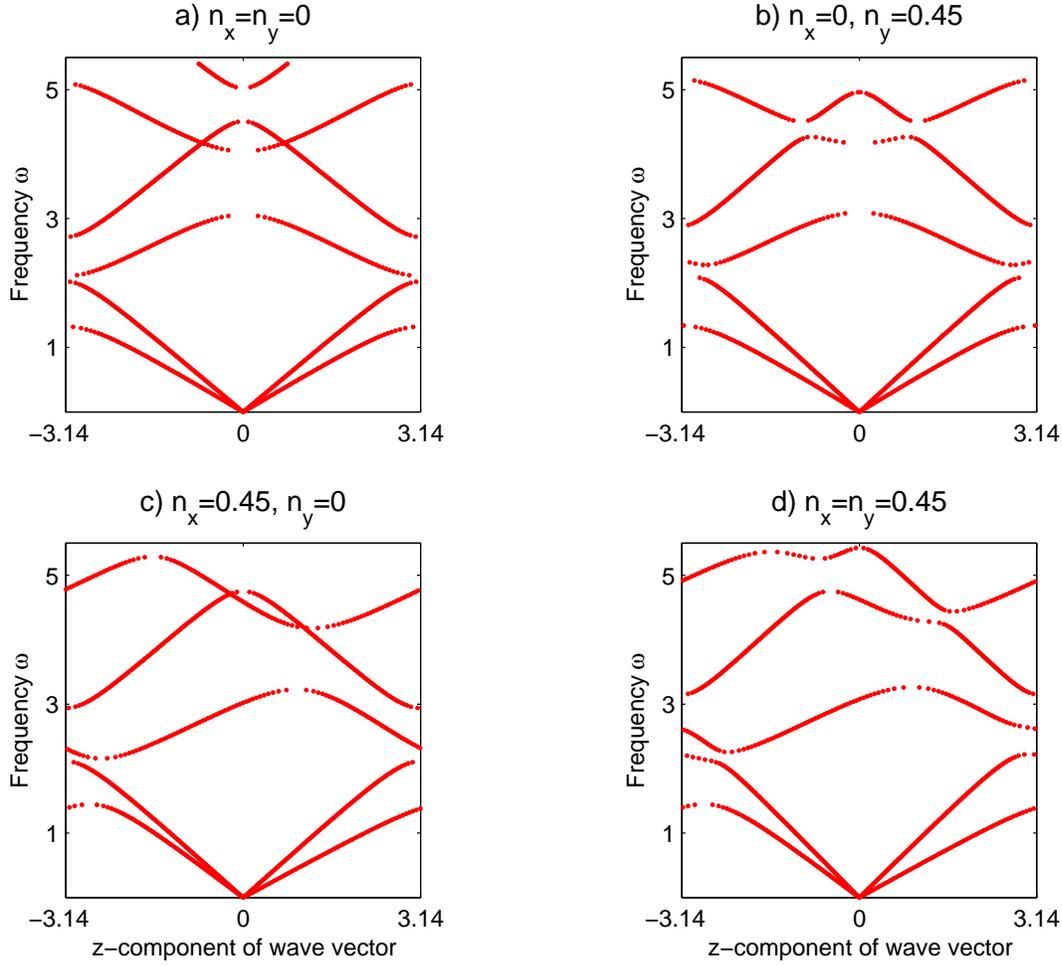}
\caption{Axial dispersion relation $\protect\omega \left( k\right) $ for
fixed $\left( n_{x},n_{y}\right) $ for the periodic array in Fig. 2. The AFM
regime can occur only if $n_{x}\neq 0$ and $n_{y}\neq 0$ (the case (d)).}
\label{DRn4}
\end{figure}

In Fig. \ref{DRn4} we show the axial dispersion relation $\omega \left(
k\right) $ of this periodic array, computed for four different directions $%
(n_{x},n_{y})$ of incident wave propagation. 
These four cases cover all the possibilities, different in terms of symmetry.

In the case (\emph{a}) of normal incidence, when $n_{x}=n_{y}=0$, the
dispersion relation is axially symmetric, as must be the case with any
reciprocal periodic stratified medium (see the explanation after Eq. (\ref%
{nrs})).

In the case (\emph{b}), when $n_{x}=0$ and $n_{y}\neq 0$, the two necessary
conditions (\ref{asymC2}) and (\ref{kxKy<>0}) for the axial spectral
asymmetry are met. Yet, those conditions prove not to be sufficient. Indeed,
if $n_{x}=0$, either of the symmetry operations%
\begin{equation}
2_{y}\text{ \ and \ \ }m_{y}^{\prime }\equiv m_{y}\times R  \label{Rm_y}
\end{equation}%
imposes the relation 
\begin{equation}
\omega \left( 0,k_{y},k\right) =\omega \left( 0,k_{y},-k\right)  \label{kx}
\end{equation}%
which implies the axial spectral symmetry. Neither stationary inflection
point, nor AFM can occur in this case.

In the case (\emph{c}), when $n_{x}\neq 0$ and $n_{y}=0$, the situation is
more complicated. The quasimomentum $\vec{k}$ lies now in the $x-z$\ plane,
which coincides with the mirror plane $m_{y}$. Therefore, every Bloch
eigenmode $\Psi _{\vec{k}}\left( z\right) $ can be classified as a pure TE
or pure TM mode, depending on the $\Psi _{\vec{k}}\left( z\right) $ parity
with respect to the mirror reflection $m_{y}$%
\begin{equation}
\text{for TE mode \ \ }m_{y}\Psi _{\vec{k}}\left( z\right) =-\Psi _{\vec{k}%
}\left( z\right) \text{; \ \ for TM mode: \ \ }m_{y}\Psi _{\vec{k}}\left(
z\right) =\Psi _{\vec{k}}\left( z\right) .  \label{parity}
\end{equation}%
The TE modes have axially symmetric dispersion relation%
\begin{equation}
\text{for TE modes: \ }\omega \left( k_{x},0,k\right) =\omega \left(
k_{x},0,-k\right) .  \label{TE}
\end{equation}%
Indeed, the component $\varepsilon _{xz}$ of the dielectric tensor $\hat{%
\varepsilon}_{A}$ does not affect the TE modes, because in this case the
electric component $\mathbf{E}(\mathbf{r},t)$ of the electromagnetic field
is parallel to the $y$ axis. As a consequence, the axial dispersion relation
of the TE spectral branches is similar to that of the isotropic case with $%
\varepsilon _{xz}=0$, where it is always symmetric. By contrast, for the TM
modes we have $\mathbf{E}(\mathbf{r},t)\perp y$. Therefore, the TM modes are
affected by $\varepsilon _{xz}$ and display axially asymmetric dispersion
relation%
\begin{equation}
\text{for TM modes: \ }\omega \left( k_{x},0,k\right) \neq \omega \left(
k_{x},0,-k\right) ,  \label{TM}
\end{equation}%
as seen in Fig. \ref{DRn4}(c). We wish to remark, though, that the equality (%
\ref{TE}) cannot be derived from symmetry arguments only. The axial spectral
symmetry of the TE modes is not exact and relies on the approximation (\ref%
{mu}) for the magnetic permeability of the $A$ layers. On the other hand,
the fact that the spectral branches have different parity (\ref{parity})
with respect to the symmetry operation$\ m_{y}$, implies that none of the
branches can develop a stationary inflection point (see Eq. (\ref{T=TT}) and
explanations thereafter). Thus, in the case $n_{y}=0$, in spite of the axial
spectral asymmetry, the AFM\ regime cannot occur either.

Finally, in the general case (\emph{d}), when $n_{x}\neq 0$ and $n_{y}\neq 0$%
, all the spectral branches display the property (\ref{asym}) of the axial
spectral asymmetry. In addition, the Bloch eigenmodes now are of the same
symmetry (i. e., belong to the same irreducible representation of the wave
vector symmetry group) and are neither TE, nor TM. This is exactly the case
when the AFM regime can be achieved at some frequencies by proper choice of
the incident angle. For instance, if we impose the equality $n_{x}=$ $n_{y}$
and change the incident angle only, it turns out that every single spectral
branch at some point develops a stationary inflection point (\ref{nInfl})
and, thereby, displays the AFM at the respective frequency. If we want the
AFM at a specified frequency $\omega _{0}$, then we will have to adjust both 
$n_{x}$ and $n_{y}$.

\section{Electrodynamics of the axially frozen mode}

\subsection{Reduced Maxwell equations}

We start with the classical Maxwell equations for time-harmonic fields in
nonconducting media%
\begin{equation}
\nabla \times \mathbf{E}\left( \vec{r}\right) =i\frac{\omega }{c}\mathbf{B}%
\left( \vec{r}\right) ,\;\nabla \times \mathbf{H}\left( \vec{r}\right) =-i%
\frac{\omega }{c}\mathbf{D}\left( \vec{r}\right) ,  \label{THME}
\end{equation}%
where%
\begin{equation}
\mathbf{D}\left( \vec{r}\right) =\hat{\varepsilon}\left( \vec{r}\right) 
\mathbf{E}\left( \vec{r}\right) ,\ \mathbf{B}\left( \vec{r}\right) =\hat{\mu}%
\left( \vec{r}\right) \mathbf{H}\left( \vec{r}\right) ,  \label{MR}
\end{equation}%
In a lossless dielectric medium, the material tensors $\hat{\varepsilon}%
\left( \vec{r}\right) $ and $\hat{\mu}\left( \vec{r}\right) $\ are
Hermitian. In a stratified medium, the tensors $\hat{\varepsilon}\left( \vec{%
r}\right) $ and $\hat{\mu}\left( \vec{r}\right) $ depend on a single
Cartesian coordinate $z$, and$\;$the Maxwell equations (\ref{THME}) can be
recast as%
\begin{equation}
\nabla \times \mathbf{E}\left( \vec{r}\right) =i\frac{\omega }{c}\hat{\mu}%
\left( z\right) \mathbf{H}\left( \vec{r}\right) ,\;\nabla \times \mathbf{H}%
\left( \vec{r}\right) =-i\frac{\omega }{c}\hat{\varepsilon}\left( z\right) 
\mathbf{E}\left( \vec{r}\right) .  \label{MEz}
\end{equation}%
Solutions for Eq. (\ref{MEz}) are sought in the following form%
\begin{equation}
\mathbf{E}\left( \vec{r}\right) =e^{i\left( k_{x}x+k_{y}y\right) }\vec{E}%
\left( z\right) ,\ \mathbf{H}\left( \vec{r}\right) =e^{i\left(
k_{x}x+k_{y}y\right) }\vec{H}\left( z\right) .  \label{LEM}
\end{equation}%
$\allowbreak $The substitution (\ref{LEM}) transforms the system of six
linear equation (\ref{MEz}) into a system of four linear differential
equations%
\begin{equation}
\partial _{z}\Psi \left( z\right) =i\frac{\omega }{c}M\left( z\right) \Psi
\left( z\right) ,\;\Psi \left( z\right) =\left[ 
\begin{array}{c}
E_{x}\left( z\right) \\ 
E_{y}\left( z\right) \\ 
H_{x}\left( z\right) \\ 
H_{y}\left( z\right)%
\end{array}%
\right]  \label{ME4}
\end{equation}%
The explicit expression for the Maxwell operator $M\left( z\right) $ is%
\begin{equation}
M\left( z\right) =\left[ 
\begin{array}{cc}
M_{11} & M_{12} \\ 
M_{21} & M_{22}%
\end{array}%
\right]  \label{M}
\end{equation}%
where%
\begin{eqnarray*}
M_{11} &=&\left[ 
\begin{array}{cc}
-\frac{\varepsilon _{xz}^{\ast }}{\varepsilon _{zz}}n_{x}-\frac{\mu _{yz}}{%
\mu _{zz}}n_{y} & \left( -\frac{\varepsilon _{yz}^{\ast }}{\varepsilon _{zz}}%
+\frac{\mu _{yz}}{\mu _{zz}}\right) n_{x} \\ 
-\left( \frac{\varepsilon _{xz}^{\ast }}{\varepsilon _{zz}}-\frac{\mu _{xz}}{%
\mu _{zz}}\right) n_{y} & -\frac{\varepsilon _{yz}^{\ast }}{\varepsilon _{zz}%
}n_{y}-\frac{\mu _{xz}}{\mu _{zz}}n_{x}%
\end{array}%
\right] , \\
M_{22} &=&\left[ 
\begin{array}{cc}
-\frac{\varepsilon _{yz}}{\varepsilon _{zz}}n_{y}-\frac{\mu _{xz}^{\ast }}{%
\mu _{zz}}n_{x} & \left( \frac{\varepsilon _{yz}}{\varepsilon _{zz}}-\frac{%
\mu _{yz}^{\ast }}{\mu _{zz}}\right) n_{x} \\ 
\left( \frac{\varepsilon _{xz}}{\varepsilon _{zz}}-\frac{\mu _{xz}^{\ast }}{%
\mu _{zz}}\right) n_{y} & -\frac{\varepsilon _{xz}}{\varepsilon _{zz}}n_{x}-%
\frac{\mu _{yz}^{\ast }}{\mu _{zz}}n_{y}%
\end{array}%
\right] , \\
M_{12} &=&\left[ 
\begin{array}{cc}
\mu _{xy}^{\ast }-\frac{\mu _{xz}^{\ast }\mu _{yz}}{\mu _{zz}}+\frac{%
n_{x}n_{y}}{\varepsilon _{zz}} & \mu _{yy}-\frac{\mu _{yz}\mu _{yz}^{\ast }}{%
\mu _{zz}}-\frac{n_{x}^{2}}{\varepsilon _{zz}} \\ 
-\mu _{xx}+\frac{\mu _{xz}\mu _{xz}^{\ast }}{\mu _{zz}}+\frac{n_{y}^{2}}{%
\varepsilon _{zz}} & -\mu _{xy}+\frac{\mu _{xz}\mu _{yz}^{\ast }}{\mu _{zz}}-%
\frac{n_{x}n_{y}}{\varepsilon _{zz}}%
\end{array}%
\right] , \\
M_{21} &=&\left[ 
\begin{array}{cc}
-\varepsilon _{xy}^{\ast }+\frac{\varepsilon _{xz}^{\ast }\varepsilon _{yz}}{%
\varepsilon _{zz}}-\frac{n_{x}n_{y}}{\mu _{zz}} & -\varepsilon _{yy}+\frac{%
\varepsilon _{yz}\varepsilon _{yz}^{\ast }}{\varepsilon _{zz}}+\frac{%
n_{x}^{2}}{\mu _{zz}} \\ 
\varepsilon _{xx}-\frac{\varepsilon _{xz}\varepsilon _{xz}^{\ast }}{%
\varepsilon _{zz}}-\frac{n_{y}^{2}}{\mu _{zz}} & \varepsilon _{xy}-\frac{%
\varepsilon _{xz}\varepsilon _{yz}^{\ast }}{\varepsilon _{zz}}+\frac{%
n_{x}n_{y}}{\mu _{zz}}%
\end{array}%
\right] .
\end{eqnarray*}%
The Cartesian components of the material tensors $\hat{\varepsilon}$ and $%
\hat{\mu}$ are functions of $z$ and (in dispersive media) $\omega $. The
reduced Maxwell equation (\ref{ME4}) should be complemented with the
following expressions for the $z$ components of the fields%
\begin{equation}
\begin{array}{c}
E_{z}=\left( -n_{x}H_{y}+n_{y}H_{x}-\varepsilon _{13}^{\ast
}E_{x}-\varepsilon _{23}^{\ast }E_{y}\right) \varepsilon _{zz}^{-1} \\ 
H_{z}=\left( n_{x}E_{y}-n_{y}E_{x}-\mu _{13}^{\ast }H_{x}-\mu _{23}^{\ast
}H_{y}\right) \mu _{zz}^{-1}%
\end{array}
\label{EzHz}
\end{equation}%
where $\left( n_{x},n_{y}\right) $ are defined in Eq. (\ref{n(k)}).

Notice that in the case of normal incidence, the Maxwell operator is
drastically simplified%
\begin{equation}
\text{ \ }M_{11}=M_{22}=0,\;\;\;\text{for \ }n_{x}=n_{y}=0\text{.}
\label{M11=M22=0}
\end{equation}%
This is the case we dealt with in \cite{PRB03} when considering the
phenomenon of electromagnetic unidirectionality in nonreciprocal magnetic
photonic crystals. By contrast, the objective of this Section is to show how
the terms $M_{11}$ and $M_{22}$, occurring only in the case of oblique
incidence, can lead to the phenomenon of AFM, regardless of whether or not
the nonreciprocal effects are present. Note that $M_{11}$ and $M_{22}$ are
also nonzero in materials with linear magnetoelectric effect (see, for
example, Ref. 15 and references therein), but we are not considering here
such an exotic situation.

Importantly, the $4\times 4$ matrix $M\left( z\right) $ in Eq. (\ref{M}) has
the property of $J$ - Hermitivity defined as%
\begin{equation}
\left( JM\right) ^{\dagger }=JM  \label{JM}
\end{equation}%
where%
\begin{equation}
J=J^{-1}=\left[ 
\begin{array}{cccc}
0 & 0 & 0 & 1 \\ 
0 & 0 & -1 & 0 \\ 
0 & -1 & 0 & 0 \\ 
1 & 0 & 0 & 0%
\end{array}%
\right]  \label{JJ=1}
\end{equation}

Different versions of the reduced Maxwell equation (\ref{ME4}) can be found
in the extensive literature on electrodynamics of stratified media (see, for
example, \cite{Tmatrix,Abdul00,Abdul99}, and references therein). For more
detailed studies of $J$ - Hermitian and $J$ - unitary operators see \cite%
{JHerm}.

\subsection{The transfer matrix}

The Cauchy problem%
\begin{equation}
\partial _{z}\Psi \left( z\right) =i\frac{\omega }{c}M\left( z\right) \Psi
\left( z\right) ,\;\Psi \left( z_{0}\right) =\Psi _{0}  \label{CauPsi}
\end{equation}%
for the reduced Maxwell equation (\ref{ME4}) has a unique solution%
\begin{equation}
\Psi \left( z\right) =T\left( z,z_{0}\right) \Psi \left( z_{0}\right)
\label{T(zz0)}
\end{equation}%
where the $4\times 4$ matrix $T\left( z,z_{0}\right) $ is so-called \emph{%
transfer matrix}. From the definition (\ref{T(zz0)}) of the transfer matrix
it follows that%
\begin{equation}
T\left( z,z_{0}\right) =T\left( z,z^{\prime }\right) T\left( z^{\prime
},z_{0}\right) ,\;T\left( z,z_{0}\right) =T^{-1}\left( z_{0},z\right)
,\;T\left( z,z\right) =I.  \label{Tzz0}
\end{equation}%
The matrix $T\left( z,z_{0}\right) $\ is uniquely defined by the following
Cauchy problem%
\begin{equation}
\partial _{z}T\left( z,z_{0}\right) =i\frac{\omega }{c}M\left( z\right)
T\left( z,z_{0}\right) ,\;T\left( z,z\right) =I.  \label{CauT}
\end{equation}%
The equation (\ref{CauT}), together with $J$ - Hermitivity (\ref{JM}) of the
Maxwell operator $M\left( z\right) $, imply that the matrix $T\left(
z,z_{0}\right) $ is $J$ - \emph{unitarity}, i.e.,%
\begin{equation}
T^{\dagger }\left( z,z_{0}\right) =JT^{-1}\left( z,z_{0}\right) J.
\label{JU}
\end{equation}%
(see the proof in Appendix 1). The $J$ - unitarity (\ref{JU}) of the
transfer matrix imposes strong constraints on its eigenvalues (see Eq. (\ref%
{k=k*})). It also implies that%
\begin{equation}
\left| \det T\left( z,z_{0}\right) \right| =1.  \label{|detT|=1}
\end{equation}

The transfer matrix $T_{S}$ of a stack of layers is a sequential product of
the transfer matrices $T_{m}$ of the constitutive layers%
\begin{equation}
T_{S}=\prod_{m}T_{m}  \label{TS}
\end{equation}%
If the individual layers are homogeneous, the corresponding single-layer
transfer matrices $T_{m}$ are explicitly expressed in terms of the
respective Maxwell operators $M_{m}$%
\begin{equation}
T_{m}=\exp \left( iz_{m}M_{m}\right)  \label{Tm}
\end{equation}%
where $z_{m}$ is the thickness of the $m$-th layer. The explicit expression
for $M_{m}$ is given by (\ref{M}). Thus, formula (\ref{TS}), together with (%
\ref{Tm}) and (\ref{M}), gives us an explicit expression for the transfer
matrix $T_{S}$ of an arbitrary stack of anisotropic dielectric layers. $%
T_{S} $ is a function of (i) the material tensors $\hat{\varepsilon}$ and $%
\hat{\mu}$ in each layer of the stack, (ii) the layer thicknesses, (iii) the
frequency $\omega $, and (iv) the tangential components $\left(
k_{x},k_{y}\right) =\left( n_{x}\omega /c,n_{y}\omega /c\right) $ of the
wave vector.

Consider the important particular case of normal wave propagation. Using Eq.
(\ref{Tm}) and the explicit expression (\ref{M}) for the Maxwell operator,
one can prove that%
\begin{equation}
\det \left( T_{S}\right) =1,\;\;\;\text{for }n_{x}=n_{y}=0.  \label{detT=1}
\end{equation}

Additional information related to the transfer matrix formalism can be found
in \cite{Tmatrix,Abdul00,Abdul99}\ and references therein.

\subsection{Periodic arrays. Bloch eigenmodes.}

In a periodic layered structure, all material tensors, along with the $J$ -
Hermitian matrix $M(z)$ in Eq. (\ref{ME4}), are periodic functions of $z$%
\begin{equation}
M\left( z+L\right) =M\left( z\right)  \label{A(z+L)}
\end{equation}%
where $L$ is the length of a primitive cell of the periodic stack. By
definition, Bloch solutions $\Psi _{k}\left( z\right) $ of the reduced
Maxwell equation (\ref{ME4}) with the periodic operator $M(z)$ satisfy%
\begin{equation}
\Psi _{k}\left( z+L\right) =e^{ikL}\Psi _{k}\left( z\right)  \label{Bloch}
\end{equation}%
The definition (\ref{T(zz0)}) of the $T$ - matrix together with Eq. (\ref%
{Bloch}) give%
\begin{equation}
\Psi _{k}\left( z+L\right) =T\left( z+L,z\right) \Psi _{k}\left( z\right)
=e^{ikL}\Psi _{k}\left( z\right) .  \label{Psi=TPsi}
\end{equation}%
Introducing the transfer matrix of a primitive cell%
\begin{equation}
T_{L}=T\left( L,0\right)  \label{TL}
\end{equation}%
we have from Eq. (\ref{Psi=TPsi})%
\begin{equation}
T_{L}\Phi _{k}=e^{ikL}\Phi _{k},\text{ \ \ where }\;\Phi _{k}=\Psi
_{k}\left( 0\right) .  \label{T(L)=e(ikL)}
\end{equation}%
Thus, the eigenvectors of the transfer matrix $T_{L}$ of the unit cell are
uniquely related to the eigenmodes of the reduced Maxwell equation (\ref{ME4}%
) through the relations%
\begin{equation}
\Phi _{k_{1}}=\Psi _{k_{1}}\left( 0\right) ,\;\Phi _{k_{2}}=\Psi
_{k_{2}}\left( 0\right) ,\;\Phi _{k_{3}}=\Psi _{k_{3}}\left( 0\right)
,\;\;\Phi _{k_{4}}=\Psi _{k_{4}}\left( 0\right)  \label{Psi 1234}
\end{equation}%
The respective four eigenvalues%
\begin{equation}
X_{i}=e^{ik_{i}L},\;i=1,2,3,4  \label{X(k)}
\end{equation}%
of $T_{L}$ are the roots of the characteristic equation%
\begin{equation}
F\left( X\right) =0,\text{ \ where \ }F\left( X\right) =\det \left( T_{L}-X%
\hat{I}\right) =X^{4}+P_{3}X^{3}+P_{2}X^{2}+P_{1}X+1.  \label{Char X}
\end{equation}%
For any given $\omega $ and $\left( k_{x},k_{y}\right) $, the characteristic
equation defines a set of four values $\left\{
X_{1},X_{2},X_{3},X_{4}\right\} $, or equivalently, $\left\{
k_{1},k_{2},k_{3},k_{4}\right\} $. Real $k$ correspond to propagating Bloch
waves (extended modes), while complex $k$ correspond to evanescent modes.
Evanescent modes are relevant near photonic crystal boundaries and other
structural irregularities.

The $J$-unitarity (\ref{JU}) of $T_{L}$ imposes the following restriction on
the eigenvalues (\ref{X(k)}) for any given $\omega $ and $\left(
k_{x},k_{y}\right) $%
\begin{equation}
\{k_{i}\}\equiv \{k_{i}^{\ast }\},\;i=1,2,3,4.  \label{k=k*}
\end{equation}%
In view of the relation (\ref{k=k*}), one has to consider three different
situation. The first possibility%
\begin{equation}
k_{1}\equiv k_{1}^{\ast },\;k_{2}\equiv k_{2}^{\ast },\;k_{3}\equiv
k_{3}^{\ast },\;k_{4}\equiv k_{4}^{\ast }  \label{4ex}
\end{equation}%
relates to the case of all four Bloch eigenmodes being extended. The second
possibility%
\begin{equation}
k_{1}=k_{1}^{\ast },\;k_{2}=k_{2}^{\ast },\;k_{3}=k_{4}^{\ast },\;\text{%
where }k_{3}\neq k_{3}^{\ast },\;k_{4}\neq k_{4}^{\ast },  \label{2ex2ev}
\end{equation}%
relates to the case of two extended and two evanescent modes. The last
possibility%
\begin{equation}
k_{1}=k_{2}^{\ast },\;k_{3}=k_{4}^{\ast },\;\text{where }k_{1}\neq
k_{1}^{\ast },\;k_{2}\neq k_{2}^{\ast },\;k_{3}\neq k_{3}^{\ast
},\;k_{4}\neq k_{4}^{\ast }  \label{4ev}
\end{equation}%
relates the case of a frequency gap, when all four Bloch eigenmodes are
evanescent.

Observe that the relation%
\begin{equation*}
k_{1}+k_{2}+k_{3}+k_{4}\equiv 0
\end{equation*}%
valid in the case of normal incidence (see Refs. \cite{PRE01,PRB03}), may
not apply now.

\subsubsection{Axial spectral symmetry}

Assume that the transfer matrix $T_{L}$ is similar to its inverse%
\begin{equation}
T_{L}=U^{-1}T_{L}^{-1}U  \label{TL -- 1/TL}
\end{equation}%
where $U$\ is an invertible $4\times 4$ matrix. This assumption together
with the property (\ref{JU}) of $J$-unitarity, imply the similarity of $%
T_{L} $ and $T_{L}^{\dagger }$%
\begin{equation}
T_{L}=V^{-1}T_{L}^{\dagger }V\text{, \ where \ \ }V=JU.  \label{TL -- TL*}
\end{equation}%
This relation imposes additional restrictions on the eigenvalues (\ref{X(k)}%
) for a given frequency $\omega $ and given $\left( k_{x},k_{y}\right) $%
\begin{equation}
\{k_{i}\}\equiv \{-k_{i}\},\;i=1,2,3,4.  \label{k=-k}
\end{equation}%
The relation (\ref{k=-k}) is referred to as the \emph{axial spectral symmetry%
}, because in terms of the corresponding axial dispersion relation, it
implies the equality (\ref{sym z}) for every spectral branch.

If the sufficient condition (\ref{TL -- 1/TL}) for the axial spectral
symmetry is not in place, then we can have for a given $\omega $ and $\left(
k_{x},k_{y}\right) $%
\begin{equation}
\{k_{i}\}\neq \{-k_{i}\},\;i=1,2,3,4  \label{k<>-k}
\end{equation}%
which implies the \emph{axial spectral asymmetry} (\ref{asym}).

\subsection{Stationary inflection point}

The coefficients of the characteristic polynomial $F\left( X\right) $ in Eq.
(\ref{Char X}) are functions of $\omega $ and $\left( k_{x},k_{y}\right) $.
Let $F_{0}\left( X\right) $ be the characteristic polynomial at the
stationary inflection point (\ref{nInfl}), where $\omega =\omega _{0}$ and $%
\left( k_{x},k_{y}\right) =\left( k_{0x},k_{0y}\right) $. The stationary
inflection point (\ref{nInfl}) can also be defined as follows%
\begin{equation}
F_{0}\left( X\right) =0,\;F_{0}^{\prime }\left( X\right) =0,\;F_{0}^{\prime
\prime }\left( X\right) =0,\;F_{0}^{\prime \prime \prime }\left( X\right)
\neq 0.  \label{F=F'=F''=0}
\end{equation}%
This relation requires the respective value of $X_{0}=\exp \left(
ik_{0}L\right) $ to be a triple root of the characteristic polynomial $%
F_{0}\left( X\right) $ implying 
\begin{equation}
F_{0}\left( X\right) =\left( X-X_{1}\right) \left( X-X_{0}\right) ^{3}=0.
\label{F0}
\end{equation}

A small deviation of the frequency $\omega $ from its critical value $\omega
_{0}$ changes the coefficients of the characteristic polynomial and removes
the triple degeneracy of the solution $X_{0}$%
\begin{equation}
X-X_{0}\thickapprox -6^{1/3}\left( \frac{\partial F_{0}/\partial \omega }{%
\partial ^{3}F_{0}/\partial X^{3}}\right) ^{1/3}\left( \omega -\omega
_{0}\right) ^{1/3}\xi ,\;\ \xi =1,e^{2\pi i/3},e^{-2\pi i/3}.  \label{X-X0}
\end{equation}%
or, in terms of the axial quasimomentum $k$%
\begin{equation}
k-k_{0}\thickapprox 6^{1/3}\left( \frac{\omega -\omega _{0}}{\omega
_{0}^{\prime \prime \prime }}\right) ^{1/3}\xi ,\;\;\xi =1,e^{2\pi
i/3},e^{-2\pi i/3}  \label{k-k0}
\end{equation}%
where%
\begin{equation}
\omega _{0}^{\prime \prime \prime }=\left. \left( \frac{\partial ^{3}\omega 
}{\partial k^{3}}\right) _{k_{x},k_{y}}\right| _{\vec{k}=\vec{k}_{0}}>0.
\label{omega'''}
\end{equation}%
The three solutions (\ref{k-k0}) can also be rearranged as%
\begin{equation}
\left\{ 
\begin{array}{c}
k_{ex}\thickapprox k_{0}+6^{1/3}\left( \omega _{0}^{\prime \prime \prime
}\right) ^{-1/3}\left( \omega -\omega _{0}\right) ^{1/3}, \\ 
k_{ev}\thickapprox k_{0}+\frac{1}{2}\left( 6\right) ^{1/3}\left( \omega
_{0}^{\prime \prime \prime }\right) ^{-1/3}\left( \omega -\omega _{0}\right)
^{1/3}+i\frac{\sqrt{3}}{2}6^{1/3}\left( \omega _{0}^{\prime \prime \prime
}\right) ^{-1/3}\left| \omega -\omega _{0}\right| ^{1/3}, \\ 
k_{EV}\thickapprox k_{0}+\frac{1}{2}\left( 6\right) ^{1/3}\left( \omega
_{0}^{\prime \prime \prime }\right) ^{-1/3}\left( \omega -\omega _{0}\right)
^{1/3}-i\frac{\sqrt{3}}{2}6^{1/3}\left( \omega _{0}^{\prime \prime \prime
}\right) ^{-1/3}\left| \omega -\omega _{0}\right| ^{1/3}.%
\end{array}%
\right.  \label{k,k,k}
\end{equation}%
The real $k_{ex}$ in (\ref{k,k,k}) relates to the extended mode $\Psi
_{ex}\left( z\right) $, with $u_{z}=0$ at $\omega =\omega _{0}$. The other
two solutions, $k_{ev}$ and $k_{EV}=k_{ev}^{\ast }$, correspond to a pair of
evanescent modes $\Psi _{ev}\left( z\right) $ and $\Psi _{EV}\left( z\right) 
$ with positive and negative infinitesimally small imaginary parts,
respectively. Those modes are truly evanescent (i.e., have $\Im k\neq 0$)
only if $\omega \neq \omega _{0}$, but it does not mean that at $\omega
=\omega _{0}$, the eigenmodes $\Psi _{ev}\left( z\right) $ and $\Psi
_{EV}\left( z\right) $ become extended. In what follows we will take a
closer look at this problem.

\subsubsection{Eigenmodes at the frequency of AFM}

Consider the vicinity of stationary inflection point (\ref{kInfl}). As long
as $\omega \neq \omega _{0}$, the four eigenvectors (\ref{Psi 1234}) of the
transfer matrix $T_{L}$ comprise two extended and two evanescent Bloch
solutions. One of the extended modes (say, $\Phi _{k_{1}}$) corresponds to
the non-degenerate real root $X_{1}=e^{ik_{1}L}$ of the characteristic
equation (\ref{Char X}). This mode has negative axial group velocity $%
u_{z}\left( k_{1}\right) <0$ and, therefor, is of no interest for us. The
other three eigenvectors of $T_{L}$ correspond to three nearly degenerate
roots (\ref{X-X0}). As $\omega $ approaches $\omega _{0}$, these three
eigenvalues become degenerate, while the respective three eigenvectors $\Phi
_{k_{2}},\Phi _{k_{3}},$ and $\Phi _{k_{4}}$ become collinear%
\begin{equation}
\Phi _{k_{2}}\rightarrow \alpha _{1}\Phi _{k_{0}},\;\Phi _{k_{3}}\rightarrow
\alpha _{2}\Phi _{k_{0}},\;\Phi _{k_{4}}\rightarrow \alpha _{3}\Phi
_{k_{0}},\;\ \ \text{as \ }\omega \rightarrow \omega _{0}.  \label{colin234}
\end{equation}%
The latter important feature relates to the fact that at $\omega =\omega
_{0} $, the matrix $T_{L}\ $has a nontrivial Jordan canonical form%
\begin{equation}
U^{-1}T_{L}U=\left[ 
\begin{array}{cccc}
X_{1} & 0 & 0 & 0 \\ 
0 & X_{0} & 1 & 0 \\ 
0 & 0 & X_{0} & 1 \\ 
0 & 0 & 0 & X_{0}%
\end{array}%
\right] ,\;\;\text{at \ }\omega =\omega _{0}  \label{TL0}
\end{equation}%
and, therefore, cannot be diagonalized. It is shown rigorously in \cite%
{PRB03}, that the very fact that the $T_{L}$ eigenvalues display the
singularity (\ref{X-X0}), implies that at \ $\omega =\omega _{0}$, the
matrix $T_{L}$ has the canonical form (\ref{TL0}). In line with (\ref%
{colin234}), the matrix $T_{L}$ from Eq. (\ref{TL0}) has only two (not
four!) eigenvectors:

\begin{enumerate}
\item $\Phi _{k_{1}}=\Psi _{k_{1}}\left( 0\right) $, corresponding to the
non-degenerate root $X_{1}$ and relating to the extended mode with $u_{z}<0$;

\item $\Phi _{k_{0}}=\Psi _{k_{0}}\left( 0\right) $, corresponding to the
triple root $X_{0}$ and related to the AFM.
\end{enumerate}

The other two solutions of the Maxwell equation (\ref{ME4}) at $\omega
=\omega _{0}$ are general Floquet eigenmodes, which do not reduce to the
canonical Bloch form (\ref{Bloch}). Yet, they can be related to $\Psi
_{k_{0}}\left( z\right) $. Indeed, following the standard procedure (see,
for example, \cite{Linalg,LODE}), consider an extended Bloch solution $\Psi
_{k}\left( z\right) $ of the reduced Maxwell equation (\ref{ME4})%
\begin{equation}
\mathbf{L}\Psi _{k}\left( z\right) =0,\;\text{where}\;\mathbf{L}=\partial
_{z}-i\frac{\omega }{c}M\left( z\right)  \label{LPsi_k}
\end{equation}%
where both operators $M\left( z\right) $ and $\mathbf{L}\left( z\right) $\
are functions of $\omega $ and $\left( k_{x},k_{y}\right) $. Assume now that
the axial dispersion relation $\omega \left( k\right) $ has a stationary
inflection point (\ref{kInfl}) at $k=k_{0}$. Differentiating Eq. (\ref%
{LPsi_k}) with respect to $k$ at constant $\left( k_{x},k_{y}\right) $
gives, with consideration for Eq. (\ref{kInfl}),%
\begin{equation*}
\mathbf{L}\partial _{k}\Psi _{k}\left( z\right) =0,\;\mathbf{L}\partial
_{kk}^{2}\Psi _{k}\left( z\right) =0,\;\;\text{at\ }k=k_{0}.
\end{equation*}%
This implies that at $k=k_{0}$, both functions%
\begin{equation}
\Psi _{01}\left( z\right) =\left. \partial _{k}\Psi _{k}\left( z\right)
\right| _{k=k_{0}},\;\text{and}\ \ \Psi _{02}\left( z\right) =\left.
\partial _{kk}^{2}\Psi _{k}\left( z\right) \right| _{k=k_{0}}  \label{Floq}
\end{equation}%
are also eigenmodes of the reduced Maxwell equation at $\omega =\omega _{0}$%
. Representing $\Psi _{k}\left( z\right) $ in the form%
\begin{equation}
\Psi _{k}\left( z\right) =\psi _{k}\left( z\right) e^{ikz},\;\text{where\ }%
\psi _{k}\left( z+L\right) =\psi _{k}\left( L\right) ,\;\Im k=0
\label{Psi_k}
\end{equation}%
and substituting Eq. (\ref{Psi_k}) into (\ref{Floq}) we get%
\begin{eqnarray}
\Psi _{01}\left( z\right) &=&\bar{\Psi}_{k_{0}}\left( z\right) +iz\Psi
_{k_{0}}\left( z\right) ,  \label{Psi 01} \\
\Psi _{02}\left( z\right) &=&\bar{\Psi}_{k_{0}}^{\prime }\left( z\right) +iz%
\bar{\Psi}_{k_{0}}\left( z\right) -z^{2}\Psi _{k_{0}}\left( z\right) ,
\label{Psi 02}
\end{eqnarray}%
where%
\begin{equation*}
\bar{\Psi}_{k_{0}}\left( z\right) =\left( \partial _{k}\psi _{k}\left(
z\right) \right) _{k=k_{0}}e^{ik_{0}z}\;\text{and }\;\bar{\Psi}%
_{k_{0}}^{\prime }\left( z\right) =\left( \partial _{kk}^{2}\psi _{k}\left(
z\right) \right) _{k=k_{0}}e^{ik_{0}z}
\end{equation*}%
are auxiliary Bloch functions (not eigenmodes).

To summarize, at the frequency $\omega _{0}$ of AFM, there are four
solutions for the reduced Maxwell equation (\ref{ME4})%
\begin{equation}
\Psi _{k_{1}}\left( z\right) ,\Psi _{k_{0}}\left( z\right) ,\Psi _{01}\left(
z\right) ,\Psi _{02}\left( z\right)  \label{4sol}
\end{equation}%
The first two solutions from (\ref{4sol}) are extended Bloch eigenmodes with 
$u_{z}<0$ and $u_{z}=0$, respectively. The other two solutions diverges as
the first and the second power of $z$, respectively, they are referred to as
general (non-Bloch) Floquet modes.

Deviation of the frequency $\omega $ from $\omega _{0}$ removes the triple
degeneracy (\ref{TL0}) of the matrix $T_{L}$, as seen from Eq. (\ref{X-X0}).
The modified matrix $T_{L}$ can now be reduced to a diagonal form with the
set (\ref{Psi 1234}) of four eigenvectors comprising two extended and two
evanescent Bloch solutions.

\subsection{Symmetry considerations}

In Section 2, we discussed the relation between the symmetry of the axial
dispersion relation of a periodic stack, and the phenomenon of AFM. At this
point we can prove that indeed, the axial spectral asymmetry (\ref{asym}) is
a necessary condition for the occurrence of the stationary inflection point
and for the AFM associated with such a point. As we have seen earlier in
this Section, the stationary inflection point relates to a triple root of
the characteristic polynomial $F(X)$ from Eq. (\ref{Char X}). Since $F(X)$
is a polynomial of the fourth degree, it cannot have a symmetric pair of
triple roots, that would have been the case for axially symmetric dispersion
relation. Hence, only asymmetric axial dispersion relation $\omega \left(
k\right) $ can display a stationary inflection point (\ref{kInfl}) or,
equivalently, (\ref{nInfl}), as shown in Fig. \ref{tE}(b). In this respect,
the situation with the AFM is somewhat similar to that of the frozen mode in
unidirectional magnetic photonic crystals \cite{PRB03}. The difference lies
in the physical nature of the phenomenon. The bulk spectral asymmetry (\ref%
{nrs}) leading to the effect of electromagnetic unidirectionality,
essentially requires the presence of nonreciprocal magnetic materials. By
contrast, the axial spectral asymmetry (\ref{asym}) along with the AFM
regime can be realized in perfectly reciprocal periodic dielectric stacks
with symmetric bulk dispersion relation (\ref{symm B}). On the other hand,
the axial spectral asymmetry essentially requires an oblique light
incidence, which is not needed for the bulk spectral asymmetry.

Another important symmetry consideration is that in the vicinity of the
stationary inflection point (\ref{kInfl}), all four Bloch eigenmodes (\ref%
{Psi 1234}) must have the same symmetry, which means that all of them \emph{%
must belong to the same one-dimensional irreducible representation of the
Bloch wave vector group}. This condition is certainly met when the direction
defined by $\left( n_{x},n_{y}\right) $ is not special in terms of symmetry.
Let us see what happens if the above condition is not in place. Consider the
situation (\ref{parity}), when at any given frequency $\omega $ and fixed $%
\left( n_{x},n_{y}\right) =\left( n_{x},0\right) $, two of the Bloch
eigenmodes are TE modes and the other two are TM modes. Note, that TE and TM
modes belong to \emph{different} one-dimensional representations of the
Bloch wave vector group. In such a case, the transfer matrix $T_{L}$ can be
reduced to the block-diagonal form%
\begin{equation*}
T_{L}=\left[ 
\begin{array}{cccc}
T_{11} & T_{12} & 0 & 0 \\ 
T_{21} & T_{22} & 0 & 0 \\ 
0 & 0 & T_{33} & T_{34} \\ 
0 & 0 & T_{43} & T_{44}%
\end{array}%
\right]
\end{equation*}%
The respective characteristic polynomial $F(X)$ degenerates into%
\begin{equation}
F(X)=F_{TE}(X)F_{TM}(X)  \label{T=TT}
\end{equation}%
where $F_{TE}(X)$ and $F_{TM}(X)$ are independent second degree polynomials
describing the TE and TM spectral branches, respectively. Obviously, in such
a situation, the transfer matrix cannot have the nontrivial canonical form
Eq. (\ref{TL0}), and the respective axial dispersion relation cannot develop
a stationary inflection point (\ref{F=F'=F''=0}), regardless of whether or
not the axial spectral asymmetry is in place.

\section{The AFM regime in \ a semi-infinite stack}

\subsection{Boundary conditions}

In vacuum (to the left of semi-infinite slab in Fig. \ref{SIS1}) the
electromagnetic field $\Psi _{V}\left( z\right) $ is a superposition of the
incident and reflected waves%
\begin{equation}
\Psi _{V}\left( z\right) =\Psi _{I}\left( z\right) +\Psi _{R}\left( z\right)
,\text{ \ \ at }\;z<0  \label{L=I+R}
\end{equation}%
At the slab boundary\ we have%
\begin{equation}
\Psi _{V}\left( 0\right) =\Psi _{I}\left( 0\right) +\Psi _{R}\left( 0\right)
=\Phi _{I}+\Phi _{R}.  \label{I,R}
\end{equation}%
where%
\begin{eqnarray}
\Phi _{I} &=&\left[ 
\begin{array}{c}
E_{I,x} \\ 
E_{I,y} \\ 
H_{I,x} \\ 
H_{I,y}%
\end{array}%
\right] =\left[ 
\begin{array}{c}
E_{I,x} \\ 
E_{I,y} \\ 
-E_{I,x}n_{x}n_{y}n_{z}^{-1}-E_{I,y}\left( 1-n_{x}^{2}\right) n_{z}^{-1} \\ 
E_{I,x}\left( 1-n_{y}^{2}\right) n_{z}^{-1}+E_{I,y}n_{x}n_{y}n_{z}^{-1}%
\end{array}%
\right] ,  \label{Phi V} \\
\Phi _{R} &=&\left[ 
\begin{array}{c}
E_{R,x} \\ 
E_{R,y} \\ 
H_{R,x} \\ 
H_{R,y}%
\end{array}%
\right] =\left[ 
\begin{array}{c}
E_{R,x} \\ 
E_{R,y} \\ 
E_{R,x}n_{x}n_{y}n_{z}^{-1}+E_{R,y}\left( 1-n_{x}^{2}\right) n_{z}^{-1} \\ 
-E_{R,x}\left( 1-n_{y}^{2}\right) n_{z}^{-1}-E_{R,y}n_{x}n_{y}n_{z}^{-1}%
\end{array}%
\right] .  \notag
\end{eqnarray}%
The complex vectors $\vec{E}_{I},\vec{H}_{I}$ and $\vec{E}_{R},\vec{H}_{R}$
are related to the actual electromagnetic field components $\mathbf{E}_{I},%
\mathbf{H}_{I}$ and $\mathbf{E}_{R},\mathbf{H}_{R}$ as 
\begin{eqnarray*}
\mathbf{E}_{I} &=&e^{i\frac{\omega }{c}\left( n_{x}x+n_{y}y\right) }\vec{E}%
_{I}\left( z\right) ,\ \mathbf{H}_{I}=e^{i\frac{\omega }{c}\left(
n_{x}x+n_{y}y\right) }\vec{H}_{I}, \\
\mathbf{E}_{R} &=&e^{i\frac{\omega }{c}\left( n_{x}x+n_{y}y\right) }\vec{E}%
_{R}\left( z\right) ,\ \mathbf{H}_{R}=e^{i\frac{\omega }{c}\left(
n_{x}x+n_{y}y\right) }\vec{H}_{R}.
\end{eqnarray*}

The transmitted wave $\Psi _{T}\left( z\right) $ inside the semi-infinite
slab is a superposition of two Bloch eigenmodes%
\begin{equation}
\Psi _{T}\left( z\right) =\Psi _{1}\left( z\right) +\Psi _{2}\left( z\right)
,\text{ \ \ at }\;z>0.  \label{T=1+2}
\end{equation}%
(in the case of a finite slab, all four eigenmodes (\ref{Psi 1234}) would
contribute to $\Psi _{T}\left( z\right) $). The eigenmodes $\Psi _{1}\left(
z\right) \ $and $\Psi _{2}\left( z\right) $ can be both extended (with $%
u_{x}>0$), one extended and one evanescent (with $u_{x}>0$ and $\Im k>0,$
respectively), or both evanescent (with $\Im k>0$), depending on which of
the three cases (\ref{4ex}), (\ref{2ex2ev}), or (\ref{4ev}) we are dealing
with. In particular, in the vicinity of the AFM (e.g., the vicinity of $%
\omega _{0}$ in Fig. \ref{tE}(b)), we always have the situation (\ref{2ex2ev}%
). Therefore, in the vicinity of AFM, $\Psi _{T}\left( z\right) $ is a
superposition of the extended eigenmode $\Psi _{ex}\left( z\right) $ with
the group velocity $u_{z}>0$, and the evanescent mode $\Psi _{ev}\left(
z\right) $ with $\Im k>0$%
\begin{equation}
\Psi _{T}\left( z\right) =\Psi _{ex}\left( z\right) +\Psi _{ev}\left(
z\right) ,\text{ \ \ at }\;z>0.  \label{T=ex+ev}
\end{equation}%
The asymptotic expressions for the respective wave vectors $k_{ex}$ and $%
k_{ev}$ in the vicinity of AFM are given in Eq. (\ref{k,k,k}).

When the frequency $\omega $ exactly coincides with the frequency $\omega
_{0}$ of the AFM, the representation (\ref{T=ex+ev}) for $\Psi _{T}\left(
z\right) $ is not valid. In such a case, according to Eq. (\ref{k,k,k}),
there is no evanescent modes at all. It turns out that at $\omega =\omega
_{0}$, the electromagnetic field inside the slab is a superposition of the
extended mode $\Psi _{k_{0}}\left( z\right) $ and the (non-Bloch) Floquet
eigenmode $\Psi _{01}\left( z\right) $ from Eq. (\ref{Psi 01})%
\begin{equation}
\Psi _{T}\left( z\right) =\Psi _{k_{0}}\left( z\right) +\Psi _{01}\left(
z\right) ,\;\;\text{at }\omega =\omega _{0}\text{ \ and \ }z>0.
\label{ex0+01}
\end{equation}%
Since the extended eigenmode $\Psi _{k_{0}}\left( z\right) $ has zero axial
group velocity $u_{z}$, it does not contribute to the axial energy flux $%
S_{z}$. By contrast, the divergent non-Bloch contribution $\Psi _{01}\left(
z\right) $ is associated with the finite axial energy flux $S_{z}>0$,
although the notion of group velocity does not apply here. The detailed
analysis is carried out in the next subsection.

Knowing the eigenmodes inside the slab and using the standard
electromagnetic boundary conditions%
\begin{equation}
\Phi _{T}=\Phi _{I}+\Phi _{R}\text{, }\;\text{where }\Phi =\Psi \left(
0\right) ,  \label{BC}
\end{equation}%
one can express the amplitude and composition of the transmitted wave $\Psi
_{T}$ and reflected wave $\Psi _{R}$, in terms of the amplitude and
polarization of the incident wave $\Psi _{I}$. This gives us the
transmittance and reflectance coefficients (\ref{t_e,r_e}) of the
semi-infinite slab, as well as the electromagnetic field distribution $\Psi
_{T}\left( z\right) $ inside the slab, as functions of the incident wave
polarization, the direction $\vec{n}$ of incidence, and the frequency $%
\omega $.

\subsection{Field amplitude inside semi-infinite slab}

In what follows we assume that $\omega $ can be arbitrarily close but not
equal to $\omega _{0}$, unless otherwise is explicitly stated. This will
allow us to treat the transmitted wave $\Psi _{T}\left( z\right) $ as a
superposition (\ref{T=ex+ev}) of one extended and one evanescent mode. Since
evanescent modes do not transfer energy in the $z$ direction, the extended
mode is solely responsible for the axial energy flux $S_{z}$%
\begin{equation}
S_{z}\left( \Psi _{T}\right) =S_{z}\left( \Psi _{ex}\right) .  \label{SzT}
\end{equation}%
According to Eq. (\ref{S=S(z)}), $S_{z}$ does not depend on $z$ and can be
expressed in terms of the semi-infinite slab transmittance $\tau $ from Eq. (%
\ref{t_e,r_e})%
\begin{equation}
S_{z}=\tau \left( \vec{S}_{I}\right) _{z}=\tau S_{I},  \label{Sz =t}
\end{equation}%
where $S_{I}=\left( \vec{S}_{I}\right) _{z}$ is the axial energy flux of the
incident wave, which is set to be unity.

The energy density $W_{ex}$ associated with the extended mode $\Psi
_{ex}\left( z\right) $ can be expressed in terms of the axial component $%
u_{z}$ of its group velocity and the axial component $S_{z}\left( \Psi
_{ex}\right) $ of the respective energy density flux%
\begin{equation}
W_{ex}=u_{z}^{-1}S_{z}\left( \Psi _{ex}\right) =\left( \frac{\partial \omega 
}{\partial k}\right) _{k_{x},k_{y}}^{-1}\tau S_{I}.  \label{Wex=S/om'}
\end{equation}%
In close proximity of the AFM frequency $\omega _{0}$, we have according to
Eq. (\ref{kInfl})%
\begin{equation}
\omega -\omega _{0}\thickapprox \frac{1}{6}\omega _{0}^{\prime \prime \prime
}\left( k-k_{0}\right) ^{3},  \label{om-om0}
\end{equation}%
where $\omega _{0}^{\prime \prime \prime }$ is defined in Eq. (\ref{omega'''}%
). Differentiating Eq. (\ref{om-om0}) with respect to $k$%
\begin{equation}
\left( \frac{\partial \omega }{\partial k}\right) _{k_{x},k_{y}}\thickapprox 
\frac{1}{2}\omega _{0}^{\prime \prime \prime }\left( k-k_{0}\right)
^{2}\thickapprox \frac{6^{2/3}}{2}\left( \omega _{0}^{\prime \prime \prime
}\right) ^{1/3}\left( \omega -\omega _{0}\right) ^{2/3},  \label{om'}
\end{equation}%
and plugging Eq. (\ref{om'}) into (\ref{Wex=S/om'}) yields%
\begin{equation}
W_{ex}\thickapprox \frac{2}{6^{2/3}}\tau S_{I}\left( \omega _{0}^{\prime
\prime \prime }\right) ^{-1/3}\left( \omega -\omega _{0}\right) ^{-2/3},
\label{Wex}
\end{equation}%
where the transmittance $\tau $ depends on the incident wave polarization,
the frequency $\omega $, and the direction of incidence $\left(
n_{x},n_{y}\right) =\left( ck_{x}/\omega ,ck_{y}/\omega \right) $. Formula (%
\ref{Wex}) implies that the energy density $W_{ex}$ and, therefore, the
amplitude $\left| \Psi _{ex}\left( z\right) \right| =\left| \Phi
_{ex}\right| $ of the extended mode inside the stack diverge in the vicinity
of the AFM regime%
\begin{equation}
\left| \Phi _{ex}\right| \thicksim \sqrt{W_{ex}}\thicksim \sqrt{\tau S_{I}}%
\left( \omega _{0}^{\prime \prime \prime }\right) ^{-1/6}\left| \omega
-\omega _{0}\right| ^{-1/3}\;\text{as}\;\omega \rightarrow \omega _{0}.
\label{Phi_ex}
\end{equation}

The divergence of the extended mode amplitude $\left| \Phi _{ex}\right| $
imposes the similar behavior on the amplitude $\left| \Psi _{ev}\left(
0\right) \right| =\left| \Phi _{ev}\right| $ of the evanescent mode at the
slab boundary. Indeed, the boundary condition (\ref{BC}) requires that the
resulting field $\Phi _{T}=\Phi _{ex}+\Phi _{ev}$ remains limited to match
the sum $\Phi _{I}+\Phi _{R}$ of the incident and reflected waves. The
relation (\ref{BC}) together with (\ref{Phi_ex}) imply that there is a
destructive interference of the extended $\Phi _{ex}$ and evanescent $\Phi
_{ev}$ modes at the stack boundary%
\begin{equation}
\Phi _{ex}\thickapprox -\Phi _{ev}\thickapprox K\sqrt{\tau S_{I}}\left(
\omega _{0}^{\prime \prime \prime }\right) ^{-1/6}\left( \omega -\omega
_{0}\right) ^{-1/3}\Phi _{k_{0}}\;\text{as}\;\omega \rightarrow \omega _{0},
\label{Phi ex = Phi ev}
\end{equation}%
Here $\Phi _{k_{0}}$ is the normalized eigenvector of $T_{L}$ in Eq. (\ref%
{TL0}); $K$ is a dimensionless parameter. The expression (\ref{Phi ex = Phi
ev}) is in compliance with the earlier made statement (\ref{colin234}) that
the column-vectors $\Phi _{ex}$ and $\Phi _{ev}$ become collinear as $\omega
\rightarrow \omega _{0}$.

\subsubsection{Space distribution of electromagnetic field in the AFM regime}

The amplitude $\left| \Psi _{ex}\left( z\right) \right| $ of the extended
Bloch eigenmode remains constant and equal to $\left| \Phi _{ex}\right| $
from (\ref{Phi_ex}), while the amplitude of the evanescent contribution to
the resulting field decays as%
\begin{equation}
\left| \Psi _{ev}\left( z\right) \right| =\left| \Phi _{ev}\right| e^{-z\Im
k_{ev}},\text{\ \ \ where }\Im k_{ev}\thickapprox \frac{\sqrt{3}}{2}%
6^{1/3}\left( \omega _{0}^{\prime \prime \prime }\right) ^{-1/3}\left|
\omega -\omega _{0}\right| ^{1/3}.  \label{Psi ev}
\end{equation}%
At $z\gg \left( \Im k_{ev}\right) ^{-1}$, the destructive interference (\ref%
{Phi ex = Phi ev}) of the extended and evanescent modes becomes ineffective,
and the only remaining contribution to $\Psi _{T}\left( z\right) $ is the
extended mode $\Psi _{ex}\left( z\right) $ with huge and independent of $z$
amplitude (\ref{Phi_ex}). This situation is graphically demonstrated in Fig. %
\ref{AMz}.

Let us see what happens when the frequency $\omega $ tends to its critical
value $\omega _{0}$. According to Eqs. (\ref{T(L)=e(ikL)}) and (\ref{T=ex+ev}%
), at $z=NL,$ $\;N=0,1,2,...$, the resulting field $\Psi _{T}\left( z\right) 
$ inside the slab can be represented as%
\begin{equation}
\Psi _{T}\left( z\right) =\Phi _{ex}e^{izk_{ex}}+\Phi _{ev}e^{izk_{ev}},
\label{Psi(N)}
\end{equation}%
Substituting $k_{ex}$ and $k_{ev}$ from Eq. (\ref{k,k,k}) in (\ref{Psi(N)}),
and taking into account the asymptotic relation (\ref{Phi ex = Phi ev}), we
have%
\begin{equation}
\Psi _{T}\left( z\right) \thickapprox \left( \Phi _{T}+zK\sqrt{\frac{\tau
S_{I}}{\omega _{0}^{\prime \prime \prime }}}6^{1/3}\left( \frac{i}{2}+\frac{%
\sqrt{3}}{2}\frac{\omega -\omega _{0}}{\left| \omega -\omega _{0}\right| }%
\right) \Phi _{k_{0}}\right) e^{izk_{0}}\;\ \text{as}\;\omega \rightarrow
\omega _{0}.  \label{PsiT0}
\end{equation}%
Although this asymptotic formula is valid only for $z=NL,$ $\;N=0,1,2,...$,
it is obviously consistent with the expression (\ref{Psi 01}) for the
non-Bloch solution $\Psi _{10}\left( z\right) $ of the Maxwell equation (\ref%
{ME4}) at $\omega =\omega _{0}$.

\subsubsection{The role of the incident wave polarization}

The incident wave polarization affects the relative contributions of the
extended and evanescent components to the resulting field $\Psi _{T}\left(
z\right) $ in Eq. (\ref{T=ex+ev}). In addition, it also affects the overall
transmittance (\ref{t_e,r_e}). The situation here is similar to that of the
normal incidence, considered in \cite{PRB03}. There are two special cases,
merging into a single one as $\omega \rightarrow \omega _{0}$. The first one
occurs when the elliptic polarization of the incident wave is chosen so that
it produces a single extended eigenmode $\Psi _{ex}\left( z\right) $ inside
the slab (no evanescent contribution to $\Psi _{T}\left( z\right) $). In
this case, $\Psi _{T}\left( z\right) $ reduces to $\Psi _{ex}\left( z\right) 
$, and its amplitude $\left| \Psi _{T}\left( z\right) \right| $ remains
limited and independent of $z$. As $\omega $ approaches $\omega _{0}$, the
respective transmittance $\tau $ vanishes in this case, and there is no AFM
regime. The second special case is when the elliptic polarization of the
incident wave is chosen so that it produces a single evanescent eigenmode $%
\Psi _{ev}\left( z\right) $ inside the slab (no extended contribution to $%
\Psi _{T}\left( z\right) $). In such a case, $\Psi _{T}\left( z\right) $
reduces to $\Psi _{ev}\left( z\right) $, and the amplitude $\left| \Psi
_{T}\left( z\right) \right| $ decays exponentially with $z$ in accordance
with Eq. (\ref{Psi ev}). The respective transmittance $\tau $ in this latter
case is zero regardless of the frequency $\omega $, because evanescent modes
do not transfer energy. Importantly, as $\omega $ approaches $\omega _{0}$,
the polarizations of the incident wave that produce either a sole extended
or a sole evanescent mode become indistinguishable, in accordance with Eq. (%
\ref{colin234}). In the vicinity of the AFM regime, the maximal
transmittance $\tau $ is achieved for the incident wave polarization
orthogonal to that exciting a single extended or evanescent eigenmode inside
the semi-infinite stack.

\subsection{Tangential energy flux}

So far we have been focusing on the axial electromagnetic field
distribution, as well as the axial energy flux $S_{z}$ inside semi-infinite
slab. At the same time, in and near the AFM regime, the overwhelmingly
stronger energy flux occurs in the tangential direction. Let us take a
closer look at this problem.

The axial energy flux $S_{z}$ is exclusively provided by the extended
contribution $\Psi _{ex}\left( z\right) $ to the resulting field $\Psi
_{T}\left( z\right) $, because the evanescent mode $\Psi _{ev}\left(
z\right) $ does not contribute to $S_{z}$. Neither $\left| \Psi _{ex}\right| 
$ nor $S_{z}$ depends on $z$ (see Eq. (\ref{S=S(z)})). By contrast, both the
extended and the evanescent modes determine the\ tangential energy flux $%
\vec{S}_{\tau }\left( z\right) $. Besides, according to Eq. (\ref{S=S(z)}),
the\ tangential energy flux depends on $z$. Far from the AFM regime, the
role of the evanescent mode is insignificant, because $\Psi _{ev}\left(
z\right) $ is appreciable only in a narrow region close to the slab
boundary. But the situation appears quite different near the AFM frequency.
Indeed, according to Eq. (\ref{Psi ev}), the imaginary part of the
respective Bloch wave vector $k_{ev}$ becomes infinitesimally small near the
critical point. As a consequence, the evanescent mode extends deep inside
the slab, so does its role in formation of $\vec{S}_{\tau }\left( z\right) $%
. The tangential energy flux $\vec{S}_{\tau }\left( z\right) $ as function
of $z$ can be directly obtained using formula (\ref{SxSy}) and the explicit
expression for $\Psi _{T}\left( z\right) =\Psi _{ex}\left( z\right) +\Psi
_{ev}\left( z\right) $. Although the explicit expression for $\vec{S}_{\tau
}\left( z\right) $ is rather complicated and cumbersome, it has very simple
and transparent structure. Indeed, the tangential energy flux can be
represented in the following form%
\begin{equation*}
\vec{S}_{\tau }\left( z\right) =\vec{u}_{\tau }W\left( z\right) ,
\end{equation*}%
where the tangential group velocity $\vec{u}_{\tau }$ behaves regularly at $%
\omega =\omega _{0}$. Therefore, the magnitude and the space distribution of
the tangential energy flux $\vec{S}_{\tau }\left( z\right) $ in and near the
AFM regime literally coincides with that of the electromagnetic energy
density $W\left( z\right) $, which is proportional to $\left| \Psi
_{T}\left( z\right) \right| ^{2}$. A typical picture of that is shown in
Fig. \ref{AMz}(a).

\section{Summary}

As we have seen in the previous Section, a distinctive characteristic of the
AFM regime is that the incident monochromatic radiation turns into a very
unusual grazing wave inside the slab, as shown schematically in Figs. \ref%
{SIS2} and \ref{Beam}. Such a grazing wave is significantly different from
that occurring in the vicinity of the total internal reflection regime,
where the transmitted wave also propagates along the interface. The most
obvious difference is that near the regime of total internal reflection, the
reflectivity approaches unity, which implies that the intensity of the
transmitted (refracted) wave vanishes. By contrast, in the case of AFM the
light reflection from the interface can be small, as shown in an example in
Fig. \ref{tE}(a). Thus, in the AFM case, a significant portion of the
incident light gets converted into the grazing wave (the AFM) with huge
amplitude, compared to that of the incident wave. For this reason, the AFM
regime can be of great utility in many applications.

Another distinctive feature of the AFM regime relates to the field
distribution inside the periodic medium. The electromagnetic field of the
AFM can be approximated by a divergent Floquet eigenmode $\Psi _{10}\left(
z\right) $ from (\ref{Psi 01}), whose magnitude $\left| \Psi _{10}\left(
z\right) \right| ^{2}$ increases as $z^2$, until nonlinear effects or other
limiting factors come into play. In fact, the field amplitude inside the
slab can exceed the amplitude of the incident plane wave by several orders
of magnitude, depending on the quality of the periodic array, the actual
number of the layers, and the width of the incident light beam.

Looking at the $z$ component of light group velocity and energy flux, we see
a dramatic slowdown of light in the vicinity of the AFM regime, with all
possible practical applications extensively discussed in the literature
(see, for example, \cite{Joann1,Scalora1,Agua01,Scalora2}, and references
therein). In principle, there can be a situation when the tangential
components $\left( u_{x},u_{y}\right) $ of the group velocity also vanish in
the AFM regime, along with the axial component $u_{z}$. Although we did not
try to achieve such a situation in our numerical experiments, it is not
prohibited and might occur if the physical parameters of the periodic array
are chosen properly. In such a case, the AFM regime reduces to its
particular case -- the frozen mode regime with $\vec{u}=0$ inside the
periodic medium. This regime would be similar to that considered in \cite%
{PRB03}, with one important difference: it is not related to the magnetic
unidirectionality and, hence, there is no need to incorporate nonreciprocal
magnetic layers in the periodic array. The latter circumstance allows to
realize the frozen mode regime at the infrared, optical, and even UV
frequency range.

\textbf{Acknowledgment and Disclaimer}. The efforts of A. Figotin and I.
Vitebskiy are sponsored by the Air Force Office of Scientific Research, Air
Force Materials Command, USAF, under grant number F49620-01-1-0567. The US
Government is authorized to reproduce and distribute reprints for
governmental purposes notwithstanding any copyright notation thereon. The
views and conclusions contained herein are those of the authors and should
not be interpreted as necessarily representing the official policies or
endorsements, either expressed or implied, of the Air Force Office of
Scientific Research or the US Government.

\section{Appendix 1. $J$-unitarity of the transfer matrix}

Let $n\times n$ matrix $T\left( z\right) $ satisfies the following Cauchy
problem%
\begin{equation}
\partial _{z}T\left( z\right) =iJA\left( z\right) T\left( z\right)
,\;T\left( 0\right) =I.  \label{TCau}
\end{equation}%
where $A\left( z\right) =JM\left( z\right) $ is a Hermitian matrix. Let us
prove that the unique solution $T\left( z\right) $ for Eq. (\ref{TCau}) is a 
$J$-unitary operator%
\begin{equation}
T^{\dagger }\left( z\right) =JT^{-1}\left( z\right) J^{-1}.  \label{pseuU}
\end{equation}%
To prove it, notice that Eq. (\ref{TCau}) implies%
\begin{equation}
\partial _{x}T^{\dagger }\left( z\right) =-T^{\dagger }\left( x\right)
iA\left( z\right) J,\;T^{\dagger }\left( 0\right) =I.  \label{TCau1}
\end{equation}%
Now, let us find the respective Cauchy problem for $T^{-1}\left( z\right) $.
Since%
\begin{equation*}
\partial _{z}\left[ T\left( z\right) T^{-1}\left( z\right) \right]
=0=T\left( z\right) \left[ \partial _{z}T^{-1}\left( z\right) \right] +\left[
\partial _{z}T\left( z\right) \right] T^{-1}\left( z\right) ,
\end{equation*}%
we have%
\begin{equation*}
\partial _{z}T^{-1}\left( z\right) =-T^{-1}\left( z\right) \left[ \partial
_{z}T\left( z\right) \right] T^{-1}\left( z\right) ,
\end{equation*}%
which in a combination with Eq. (\ref{TCau}) yields 
\begin{equation}
\partial _{z}T^{-1}\left( z\right) =-T^{-1}\left( z\right) iJA\left(
z\right) ,\;\,T^{-1}\left( 0\right) =I.  \label{TCau2}
\end{equation}%
Finally, multiplying both sides of the equality (\ref{TCau2}) by $J$ and
using the fact that $J^{2}=I$, we get the following Cauchy problem for $%
JT^{-1}\left( z\right) J$%
\begin{equation}
\partial _{z}\left[ JT^{-1}\left( z\right) J\right] =-\left[ JT^{-1}\left(
z\right) J\right] iA\left( z\right) J,\;\,JT^{-1}\left( 0\right) J=I.
\label{TCau3}
\end{equation}%
which is identical to that for $T^{\dagger }\left( z\right) $ from Eq. (\ref%
{TCau1}). Since both Cauchy problems (\ref{TCau1}) and (\ref{TCau3}) have
unique solutions, their similarity implies the relation (\ref{pseuU}) of $J$%
-unitarity.

\section{Appendix 2. Energy density flux}

The real-valued Poynting vector is defined by%
\begin{equation}
\mathbf{S}\left( \vec{r}\right) =\frac{c}{8\pi }\Re \left[ \mathbf{E}^{\ast
}\left( \vec{r}\right) \times \mathbf{H}\left( \vec{r}\right) \right] .
\label{Pt}
\end{equation}%
Substituting the representation (\ref{LEM}) for $\mathbf{E}\left( \vec{r}%
\right) $ and $\mathbf{H}\left( \vec{r}\right) $ in Eq. (\ref{Pt}) yields%
\begin{equation}
\mathbf{S}\left( \vec{r}\right) =\mathbf{S}\left( z\right) =\frac{c}{8\pi }%
\Re \left[ \vec{E}^{\ast }\left( z\right) \times \vec{H}\left( z\right) %
\right]  \label{S=[EH]}
\end{equation}%
implying that none of the three Cartesian components of the energy density
flux $\mathbf{S}$ depends on the transverse coordinates $x$\ and $y$. Energy
conservation argument implies that the\ component $S_{z}$ of the energy flux
does not depend on the coordinate $z$ either, while the transverse
components $S_{x}$\ and $S_{y}$ may depend on $z$. Indeed, in the case of
steady-state oscillations in a lossless medium we have, with consideration
for Eq. (\ref{S=[EH]})%
\begin{equation*}
\nabla \cdot \mathbf{S}=\partial _{z}S_{z}\left( z\right) =0
\end{equation*}%
which together with Eq. (\ref{S=[EH]}) gives%
\begin{equation}
S_{z}\left( \vec{r}\right) =S_{z}=const,\;S_{x}\left( \vec{r}\right)
=S_{x}\left( z\right) ,\;S_{y}\left( \vec{r}\right) =S_{y}\left( z\right) .
\label{S=S(z)}
\end{equation}%
The explicit expression for the $z$ component of the energy flux (\ref%
{S=[EH]}) is%
\begin{equation}
S_{z}=\frac{1}{2}\left[ E_{x}^{\ast }H_{y}-E_{y}^{\ast
}H_{x}+E_{x}H_{y}^{\ast }-E_{y}H_{x}^{\ast }\right] =\frac{1}{2}\left( \Psi
,J\Psi \right) .  \label{Sz(Psi)}
\end{equation}%
The tangential components of the energy flux can also be expressed in terms
of the column vector $\Psi \left( z\right) $ from Eq. (\ref{ME4}). Using the
expressions (\ref{EzHz}) for $E_{z}$\ and $H_{z}$ and eliminating these
field components from $\mathbf{S}\left( z\right) $ in Eq. (\ref{S=[EH]})
yields%
\begin{equation}
S_{x}=\frac{1}{2}\left( \Psi ,\hat{G}_{x}\Psi \right) ,\;S_{y}=\frac{1}{2}%
\left( \Psi ,\hat{G}_{y}\Psi \right) ,  \label{SxSy}
\end{equation}%
where $G_{x}$ and $G_{y}$ are Hermitian matrices%
\begin{equation*}
G_{x}=\left[ 
\begin{array}{cccc}
0 & -\frac{n_{y}}{\mu _{33}} & 0 & \frac{\varepsilon _{13}}{\varepsilon _{33}%
} \\ 
-\frac{n_{y}}{\mu _{33}} & 2\frac{n_{x}}{\mu _{33}} & -\frac{\mu _{13}^{\ast
}}{\mu _{33}} & \frac{\varepsilon _{23}}{\varepsilon _{33}}-\frac{\mu
_{23}^{\ast }}{\mu _{33}} \\ 
0 & -\frac{\mu _{13}}{\mu _{33}} & 0 & -\frac{n_{y}}{\varepsilon _{33}} \\ 
\frac{\varepsilon _{13}^{\ast }}{\varepsilon _{33}} & \frac{\varepsilon
_{23}^{\ast }}{\varepsilon _{33}}-\frac{\mu _{23}}{\mu _{33}} & -\frac{n_{y}%
}{\varepsilon _{33}} & 2\frac{n_{x}}{\varepsilon _{33}}%
\end{array}%
\right] ,G_{y}=\left[ 
\begin{array}{cccc}
2\frac{n_{y}}{\mu _{33}} & -\frac{n_{x}}{\mu _{33}} & -\frac{\varepsilon
_{13}}{\varepsilon _{33}}+\frac{\mu _{13}^{\ast }}{\mu _{33}} & \frac{\mu
_{23}^{\ast }}{\mu _{33}} \\ 
-\frac{n_{x}}{\mu _{33}} & 0 & -\frac{\varepsilon _{23}}{\varepsilon _{33}}
& 0 \\ 
-\frac{\varepsilon _{13}^{\ast }}{\varepsilon _{33}}+\frac{\mu _{13}}{\mu
_{33}} & -\frac{\varepsilon _{23}^{\ast }}{\varepsilon _{33}} & 2\frac{n_{y}%
}{\varepsilon _{33}} & -\frac{n_{x}}{\varepsilon _{33}} \\ 
\frac{\mu _{23}}{\mu _{33}} & 0 & -\frac{n_{x}}{\varepsilon _{33}} & 0%
\end{array}%
\right] .
\end{equation*}%
Both $G_{x}$ and $G_{y}$ are functions of the Cartesian coordinate $z$,
frequency $\omega $, and the direction $\vec{n}$ of incident wave
propagation.

\end{document}